\documentclass[a4paper,fleqn,usenatbib,natbib]{mnras}

\usepackage{xspace}
\usepackage{psfrag}
\usepackage{epsfig}
\usepackage{amssymb}
\usepackage{amsmath}
\usepackage[T1]{fontenc}

\newcommand{\be}{\begin{equation}}
\newcommand{\e}{\end{equation}}
\newcommand{\bear}{\begin{eqnarray}}
\newcommand{\ear}{\end{eqnarray}}

\def\cthree{\textsc{cubep}$^3$\textsc{m}\xspace}
\def\Msol{M_{\sun}}
\def\ud{\mathrm{d}}

\def\xi{x^{i}_{\ion{H}{i}}\,}
\def\xh1{x_{\ion{H}{i}\,}}
\def\xb{\bar{x}_{\ion{H}{i}}}
\def\Ph1{P_{\ion{H}{i}}}
\def\eh1{\eta_{\ion{H}{i}}}

\def\aap{AAP}

\def\apjs{ApJS}
\def\apjl{ApJL}

\def\HI{\ion{H}{i}\xspace}
\def\HII{\ion{H}{ii}\xspace}
\def\P{\hat{P}}
\def\Pl{{\mathcal P}}

\def\rhohi{\rho_{\ion{H}{i}}}
\def\rhoh{\rho_{\mathrm{M}}}
\def\Ps{P_2^{\mathrm{s}}}

\author[Majumdar et al.]{Suman Majumdar$^{1,2}$\thanks{smaju@astro.su.se},
Hannes Jensen$^{1,3}$\thanks{hjens@astro.su.se}, Garrelt Mellema$^{1}$, Emma Chapman$^{4,2}$, \newauthor
Filipe B.\ Abdalla$^{4,5}$, Kai-Yan Lee$^{1}$, Ilian T.\ Iliev$^{6}$, Keri L.\ Dixon$^{6}$, Kanan K.\ Datta$^{7}$, \newauthor
Benedetta Ciardi$^{8}$, Elizabeth R.\ Fernandez$^{9}$, Vibor Jeli{\'c}$^{9,10,11}$, \newauthor
L{\'e}on V.\ E.\ Koopmans$^{9}$, Saleem Zaroubi$^{9}$\\
$^{1}$Department of Astronomy and Oskar Klein Centre, Stockholm
University, AlbaNova, SE-10691 Stockholm, Sweden \\
$^{2}$Department of Physics, Blackett Laboratory, Imperial College, London
SW7 2AZ, UK \\
$^{3}$Department of Physics and Astronomy, Uppsala University, Uppsala, Sweden \\
$^{4}$Department of Physics and Astronomy, University College London, Gower Street, London, WC1E 6BT, U.\ K. \\
$^{5}$Department of Physics and Electronics, Rhodes University, PO Box 94, Grahamstown, 6140, South Africa\\
$^{6}$Astronomy Centre, Department of Physics \& Astronomy, Pevensey II Building, University of Sussex, Falmer, Brighton BN1 9QH, UK \\
$^{7}$Department of Physics, Presidency University, 86/1 College Street, Kolkata-700073, India \\
$^{8}$Max-Planck-Institut f{\"u}r Astrophysik, Karl-Schwarzschild-Strasse 1, D-85748 Garching b. M{\"u}nchen, Germany \\
$^{9}$Kapteyn Astronomical Institute, University of Groningen, PO Box 800, NL-9700 AV Groningen, the Netherlands \\
$^{10}$ASTRON - the Netherlands Institute for Radio Astronomy, PO Box 2, 7990 AA Dwingeloo, the Netherlands \\
$^{11}$Ru{\dj}er Bo\v{s}kovi\'{c} Institute, Bijeni\v{c}ka cesta 54, 10000 Zagreb, Croatia \\
}

\begin{document}
\title[The EoR sources \& 21-cm redshift-space distortions]{Effects of the sources of reionization on 21-cm redshift-space distortions}

\maketitle

\begin{abstract}
  The observed 21-cm signal from the epoch of reionization will be
  distorted along the line-of-sight by the peculiar velocities of
  matter particles. These \emph{redshift-space distortions} will
  affect the contrast in the signal and will also make it
  anisotropic. This anisotropy contains information about the
  cross-correlation between the matter density field and the neutral
  hydrogen field, and could thus potentially be used to extract
  information about the sources of reionization. In this paper, we
  study a collection of simulated reionization scenarios assuming
  different models for the sources of reionization. We show that the
  21-cm anisotropy is best measured by the quadrupole moment of the
  power spectrum. We find that, unless the properties of the
  reionization sources are extreme in some way, the quadrupole moment
  evolves very predictably as a function of global neutral
  fraction. This predictability implies that redshift-space
  distortions are not a very sensitive tool for distinguishing between
  reionization sources. However, the quadrupole moment can be used as
  a model-independent probe for constraining the reionization
  history. We show that such measurements can be done to some extent
  by first-generation instruments such as LOFAR, while the SKA should
  be able to measure the reionization history using the quadrupole
  moment of the power spectrum to great accuracy.
\end{abstract}


\section{Introduction}
\label{sec:intro}
One of the periods in the history of our Universe about which we know
the least is the epoch of reionization (EoR). During this epoch the
first sources of light formed and gradually ionized the neutral
hydrogen (\HI) in the intergalactic medium (IGM). Our present
understanding of this epoch is mainly constrained by observations of
the cosmic microwave background radiation (CMBR)
\citep{komatsu11,planck15} and the absorption spectra of high redshift
quasars \citep{becker01,fan03,white03,goto11,becker15}. These
observations suggest that reionization was an extended process,
spanning over the redshift range $6 \lesssim z \lesssim 15$ (see
e.g. \citealt{alvarez06,mitra13,mitra15,robertson15,bouwens15}). Such
indirect observations are, however, limited in their ability to answer
several important questions regarding the EoR. These unresolved issues
include the precise duration and timing of reionization, the
properties of major ionizing sources, the relative contribution to the
ionizing photon budget from various kinds of sources, and the typical
size and distribution of ionized bubbles.

Observations of the redshifted 21-cm line, originating
from spin flip transitions in neutral hydrogen atoms, are expected
to be the key to resolving many of these long standing issues. The
brightness temperature of the redshifted 21-cm line directly probes
the \HI distribution at the epoch where the radiation originated. Observing this
line enables us, in principle, to track the entire reionization history as it
proceeds with redshift.

Motivated by this, a huge effort is underway to detect the redshifted
21-cm signal from the EoR using low frequency radio interferometers,
such as the GMRT \citep{paciga13}, LOFAR
\citep{yatawatta13,haarlem13,jelic14}, MWA \citep{tingay13,bowman13},
PAPER \citep{parsons14} and 21CMA \citep{wang2013}. These observations
are complicated to a large degree by foreground emissions, which can
be $\sim\!4-5$ orders of magnitude stronger than the expected signal
(e.g.\ \citealt{dimatteo02,ali08,jelic08}), and system noise
\citep{morales05,mcquinn06}. So far only weak upper limits on the
21-cm signal have been obtained
\citep{paciga13,dillon14,parsons14,ali15}.

Owing to the low sensitivity of the first generation interferometers,
they will probably not be capable of directly imaging the \HI
distribution. This will have to wait for the arrival of the extremely
sensitive next generation of telescopes such as the SKA
\citep{mellema13,mellema15,koopmans15}. The first generation
telescopes are instead expected to detect and characterize the signal
through statistical estimators such as the variance (e.g.\
\citealt{patil14}) and the power spectrum (e.g.\ \citealt{pober14}).

Many studies to date have focused on the spherically averaged 21-cm
power spectrum (e.g.\
\citealt{mcquinn07,lidz08,barkana09,iliev12}). By averaging in
spherical shells in Fourier space, one can obtain good signal-to-noise
for the power spectrum, while still preserving many important features
of the signal. However, the spherically averaged power spectrum does
not contain all information about the underlying 21-cm field. For
example, as reionization progresses, the fluctuations in the \HI field
will make the 21-cm signal highly non-Gaussian. This non-Gaussianity
can not be captured by the power spectrum, but requires higher-order
statistics
\citep{bharadwaj05a,mellema06,watkinson14,mondal15b}.

Another important effect that can not be captured fully by the
spherically averaged power spectrum is the effect of \emph{redshift-space 
distortions} caused by the peculiar velocities of matter. The
coherent inflows of matter into overdense regions and the outflows of
matter from underdense regions will produce an additional red- or
blueshift in the 21-cm signal on top of the cosmological
redshift, changing the contrast of the 21-cm signal, and 
making it anisotropic
\citep{bharadwaj04,barkana05,mcquinn06,mao12,majumdar13,jensen13}.
The redshift-space distortions in the 21-cm signal have previously
generated interest due to the possibility of extracting the purely
cosmological matter power spectrum
\citep{barkana05,mcquinn06,shapiro13}. However, they also carry
interesting astrophysical information.

It has been shown in previous studies
\citep{majumdar13,jensen13,ghara14} that the anisotropy in the 21-cm
signal due to the redshift-space distortions will depend on the
topology of reionization, or in other words on the properties of the
sources of reionization. In this paper, we explore the prospects of
using the 21-cm redshift-space distortions to obtain information about
the sources of reionization. We calculate the evolution of the power
spectrum anisotropy for a collection of simulated reionization
scenarios assuming different properties of the sources of ionizing
photons. We also investigate different ways of quantifying this
anisotropy, and study the prospects of distinguishing between the
source models in observations with current and upcoming
interferometers.

The structure of this paper is as follows. In Section 2, we describe
the simulations and source models that we have used to generate a
collection of reionization scenarios. Here, we also describe the method
that we use to include the redshift-space distortions in the simulated
signal. In Section 3, we evaluate different methods of quantifying the
power spectrum anisotropy and discuss the interpretation of
the anisotropy using the quasi-linear model of \cite{mao12}. In
Section 4, we describe the observed large scale features of the
redshift-space anisotropy for the different reionization scenarios
that we have considered. In Section 5, we discuss the feasibility of
observing these anisotropic signatures in present and future
radio interferometric surveys, and what we can learn from the redshift- space 
distortions about the reionization history. Finally, in Section
6 we summarise our findings.

Throughout the paper we present our results for the cosmological
parameters from WMAP five year data release $h = 0.7$, $\Omega_{\mathrm{m}} =
0.27$, $\Omega_{\Lambda} = 0.73$, $\Omega_{\mathrm{b}} h^2 = 0.0226$
\citep{komatsu09}.

\section{Reionization simulations for different source models}
\label{sec:sim}
To study the effects of the sources of reionization on the
redshift-space anisotropy of the 21-cm signal, we simulate a number of
different reionization scenarios, assuming different source
properties. All of these simulations are based on a single $N$-body
simulation of the evolving dark matter density field.

\subsection{$N$-body simulations}
\label{sec:nbody}
The $N$-body simulations were carried out as part of the
\textsc{PRACE4LOFAR} project (\textsc{PRACE} projects 2012061089 and
2014102339) with the \cthree code \citep{harnoisderaps13}, which is
based on the older code \textsc{pmfast} \citep{merz05}. Gravitational
forces are calculated on a particle-particle basis at close distances,
and on a mesh for longer distances. The size of our simulation volume
was $500/h=714$ Mpc (comoving) along each side. We used 6912$^3$
particles of mass $4.0 \times 10^7\,{\rm \Msol}$ on a 13824$^3$ mesh,
which was then down-sampled to a 600$^3$ grid for modelling the
reionization. Further details of the $N$-body simulation can be found
in Dixon et al. (2015) (in preparation).

For each redshift output of the $N$-body simulation, haloes were
identified using a spherical overdensity scheme. The minimum halo mass
that we have used for our reionization simulations is $2.02 \times
10^9\,{\rm \Msol}$.

\subsection{Reionization simulations}
\label{sec:reion}
To generate the ionization maps for our different reionization
scenarios, we used a modified version of the semi-numerical code
described in \citet{choudhury09b,majumdar13,majumdar14}. This method
has been tested extensively against a radiative transfer simulation
and it is expected to generate the 21-cm signal from the EoR with an
accuracy of $\ge 90\%$ \citep{majumdar14} for the length scales that
we will deal with in this paper.

Like most other semi-numerical methods for simulating the EoR 21-cm
signal, our scheme is based on the excursion-set formalism developed
by \citet{furlanetto04b}, making it similar to the methods
described by \citet{zahn07,mesinger07,santos10}. In this method, one
compares the average number of photons in a specific volume with the
average number of neutral hydrogen atoms in that volume. Here we
assume that the neutral hydrogen follows the dark matter distribution.

Once we generate the dark matter density field at a fixed redshift
using the $N$-body simulation, then depending on the source model of
the reionization scenario under consideration (that we will discuss
later in this section), we also generate an instantaneous ionizing
photon field at the same redshift in a grid of the same size as that
of the dark matter density field. These density fields of photons and
neutral hydrogen are generally constructed in a grid coarser than the
actual $N$-body resolution. In our case the resolution of this grid is
$1.19\, {\rm Mpc}$, i.e.\ 600$^3$ cells.

Next, to determine the ionization state of a grid cell, we compare the
average number density of ionizing photons and neutral hydrogen atoms
around it within a spherical volume. The radius of this smoothing
sphere is then gradually increased, starting from the grid cell size
and going up to the assumed mean free path{\footnote{The mean free
    path of ionizing photons at high redshifts is largely an unknown
    quantity till date. We have used a fixed maximum smoothing radius
    of $70$ comoving Mpc at all redshifts for our fiducial
    reionization sources (i.e.\ UV sources), which is consistent with
    the findings of \citet{songaila10} at $z \sim 6$.}} of the photons
at that redshift. If, for any radius of this smoothing sphere, the
average number density of photons becomes greater than or equal to the
average number density of the neutral hydrogen, the cell
is flagged as ionized. The same procedure is repeated for all the grid
cells and an ionization field at that specific redshift is
generated. A more detailed description of this simulation method can
be found in \citet{choudhury09b} and \citet{majumdar14}.

\subsubsection{Source Models}
\label{sec:source}
Each of our simulated reionization scenarios consists of a different
combination of various sources of ionizing photons. Here, we describe
these different source types, and in Section \ref{sec:scenarios} we
describe the reionization scenarios.

\begin{enumerate}
\item {\bf Ultraviolet photons from galaxies (UV):}

  In most reionization models, galaxies residing in the collapsed dark
  matter halos are assumed to be the major sources of ionizing
  photons. To date, much is unknown about these high-redshift galaxies
  and the characteristics of their radiation. Thus most simulations
  assume that the total number of ionizing photons contributed by a
  halo of mass $M_{{\rm h}}$ which is hosting such galaxies is simply:
\begin{equation}
  N_{\gamma}(M_{{\rm h}}) = N_{{\rm ion}} \frac{M_{{\rm h}} \Omega_{{\rm b}}}{m_{{\rm p}} \Omega_{{\rm m}}}.
\label{eq:s1}
\end{equation}
Here, $N_{{\rm ion}}$ is a dimensionless constant which effectively
represents the number of photons entering in the IGM per baryon in
collapsed objects and $m_{{\rm p}}$ is the mass of a proton or
hydrogen atom. Reionization simulations (whether radiative transfer or
semi-numerical) which adopt this kind of model for the production of
the major portions of their ionizing photons generally produce a
global ``inside-out'' reionization scenario (see e.g.\
\citealt{mellema06,zahn07,choudhury09b,mesinger11} etc.).
  
We assume that all ionizing photons generated by these kinds of
sources{\footnote{Note that in our reionization scenarios we allow
    only halos of mass $\ge 2.02 \times 10^9\,{\rm \Msol}$ to host UV
    photon sources. In reality faint galaxies hosted by low mass halos
    ($10^5 \le M_{{\rm h}} \le 10^9\,{\rm \Msol}$) may produce
    significant amounts of UV photons during the EoR. However, there
    is a possibility that star formation in these low mass halos may
    also get suppressed once they have been ionized and heated to
    temperatures of $\sim 10^4$ K (see
    e.g. \citealt{couchman86,gnedin00b,dijkstra04,okamoto08}). This
    poses an uncertainty in their actual role during the EoR. Further,
    even when one includes them in the simulations, it is very
    unlikely that the resulting ionization and 21-cm topology will
    depart from its global ``inside-out'' nature, and the anisotropy
    of the 21-cm power spectrum---which is the focus of this
    paper---will not be severely sensitive to the presence of
    low mass sources (we discuss this in further details in Appendix
    \ref{sec:appendix_b}).}}  are in the ultraviolet part of the
spectrum. Thus, they only affect the IGM locally, up to a distance
limited by their mean free path.

\item {\bf Uniform ionizing background (UIB):} 

  The observed population of galaxies at high redshifts, seems unable
  to keep the universe ionized, unless there is a significant increase
  in the escape fraction of the ionizing photons from them with the
  increasing redshift or the galaxies below the detection thresholds
  of the present day surveys contribute a significant fraction of the
  total ionizing photons \citep{kuhlen12, mitra13}. An alternative
  possibility is that, if sources of hard X-ray photons (such as
  active galactic nuclei or X-ray binaries) were common in the early
  Universe, these could give rise to a more or less uniform ionizing
  background. Hard X-rays would easily escape their host galaxies and
  travel long distances before ionizing hydrogen
  \citep{mcquinn12,mesinger13}.

  We model this type of source as a completely uniform ionizing
  background that provides the same number of ionizing photons at
  every location.  The extreme case of a $100\%$ contribution of the
  ionizing photons from this kind of background would lead to a global
  ``outside-in'' reionization.

\item {\bf Soft X-ray photons (SXR):} 

  The escape fraction of extreme ultraviolet photons from their
  host galaxies is a hotly debated issue. Soft X-ray photons, on the
  other hand, would have little difficulty escaping from their host
  sources into the IGM. Also, there is a possibility that the X-ray
  production could have been more prevalent in the high redshift
  galaxies than their low redshift counterparts
  \citep{mirabel11,fragos13}.
  
  When including this type of photons in our simulations we have
  considered them to be uniformly distributed around their source
  halos within a radius equal to their mean free path at that
  redshift. To estimate the mean free path of these soft X-ray photons
  we have used equation (1) in \citet{mcquinn12} which is
  dependent on the redshift of their origin and the frequency of the
  photon. For simplicity, we assume that all the halos that we have
  identified as sources will produce soft X-ray photons and all of
  these photons will have the same energy ($200$ eV). A significant
  contribution by these soft X-ray photons may lead to a more
  homogeneous reionzation scenario than the case when the major
  portion of photons are in the ultraviolet.

\item {\bf Power law mass dependent efficiency (PL):} 
	
  For source type (i) we have assumed that the number of UV photons
  generated by a galaxy is proportional to its host halo mass. Instead
  of this it can also be assumed that the number of photons
  contributed by a halo follows a power law $N_{\gamma}(M_{{\rm h}})
  \propto M_{{\rm h}}^n$.

  We consider two cases, where the power law index is equal to $2$ and
  $3$. With this source model, higher-mass haloes produce relatively
  more ionizing photons, giving rise to fewer, but larger ionized
  regions. This model provides a crude approximation of a situation in
  which reionization is powered by rare, bright sources, such as
  quasars.  While the number density of quasars at high redshifts is
  not expected to be high enough to drive reionization (see
  e.g.\ \citealt{madau98}), this model serves as an extreme
  illustration of the type of topology expected from rare ionizing
  sources.
  
\end{enumerate}

\subsubsection{Reionization scenarios}
\label{sec:scenarios}
\begin{figure}
\psfrag{xh1}[c][c][1][0]{{\huge $\xb$}}
\psfrag{z}[c][c][1][0]{{\huge $z$}}
\includegraphics[width=0.5\textwidth,angle=0]{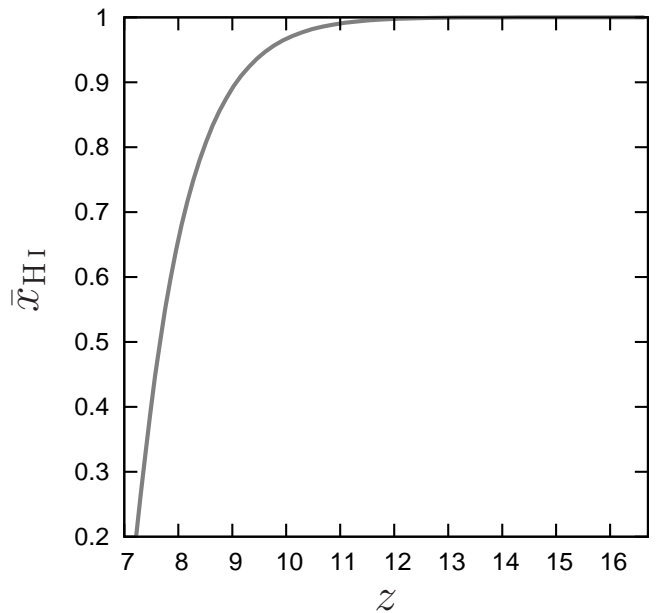}
\caption{The evolution of the mass averaged neutral fraction with
  redshift for our fiducial reionization scenario. We tune all other
  reionization scenarios to follow the same reionization history.}
\label{fig:xh1mz}
\end{figure}
The aim of this paper is to study the effect of the different source
types described above on the 21-cm power spectrum anisotropy. To do
this, we construct several reionization scenarios by combining the
source types in different ways, as summarized in Table
\ref{tab:reion}.

In our fiducial reionization scenario, $100\%$ of the ionizing photons
come from galaxies residing in dark matter halos of mass $\ge 2.02
\times 10^9 \,\Msol$, resulting in a global inside-out reionization
topology. The evolution of the reionization source population in this
scenario thus follows the evolution of the collapsed
fraction. Reionization starts around $z \sim 16$ (when the first
ionization sources formed) and by $z=7.221$ the mass averaged neutral
fraction (designated by $\xb$ throughout this paper) of the IGM
reaches $\sim 0.2$, as shown in Figure \ref{fig:xh1mz}. At this point,
the 21-cm signal is likely too weak to detect, at least with
first-generation interferometers
\citep{jensen13,majumdar13,datta14,majumdar14}.

We tune $N_{{\rm ion}}$ at each redshift for all our reionization
scenarios to follow the same evolution of $\xb$ with $z$ as the
fiducial model (see Figure \ref{fig:xh1mz}). This ensures that any
differences in the 21-cm signal and its anisotropy across different
reionization scenarios are due solely to the source types, and not due
to the underlying matter distribution.  Of course this will force all
scenarios except the fiducial one to follow an artificial reionization
history. However, our aim here is not to produce the most realistic
reionization scenarios, but rather to produce a wide range of
reionization topologies and thus a significant difference in the
anisotropy of the 21-cm signal.
\begin{table}
\centering
\begin{tabular}{c|c|c|c|c|c}
  \hline
  \hline
  Reionization & UV & UIB & SXR & PL & Non-uniform\\
  scenario & & & & $n$ & recombination \\
  \hline
  Fiducial & $100\%$ & -- & -- & $1.0$ & No \\
  Clumping & $100\%$ & -- & -- & $1.0$ & Yes \\
  UIB dominated & $20\%$ & $80\%$ & -- & $1.0$ & No \\
  SXR dominated & $20\%$ & -- & $80\%$ & $1.0$ & No \\
  UV+SXR+UIB & $50\%$ & $10\%$ & $40\%$ & $1.0$ & No \\
  PL 2.0 & -- & -- & -- & $2.0$ & No \\
  PL 3.0 & -- & -- & -- & $3.0$ & No \\
  \hline
  \hline
\end{tabular}
\caption{The relative contribution from different source types in our reionization scenarios. UV=Ultra-violet; UIB=Uniform ionizing background; SXR=Soft X-ray background; PL=Power-law.}
\label{tab:reion}
\end{table}

In all of our reionization scenarios except one, we have assumed that
the rate of recombination is uniform everywhere in the IGM. However,
in reality, the recombination rate is expected to be dependent on the
density of the ionized medium. Specifically, dense structures with
sizes on the order of a few kpc, mostly unresolvable in this type of
simulations, are expected to boost the recombination rate
significantly and thus increase the number of ionizing photons
required to complete reionization. They will also give rise to
self-shielded regions like Lyman limit systems. There has been some
effort in including these effects in the simulations of the large
scale EoR 21-cm signal (\citealt{choudhury09b,sobacchi14,choudhury14};
Shukla et al. in prep). We have included the effects of these
shelf-shielded regions in the IGM by using equation (15) in
\citet{choudhury09b}. Although this approach somewhat overestimates
the impact of non-uniform recombinations (due to the coarse resolution
of the density fields that we use in our semi-numerical models) it can
still serve to illustrate their effect on the reionization topology.

\begin{figure*}
\psfrag{Mpc}[c][c][1][0]{{\Large Mpc}}
\psfrag{mK}[c][c][1][0]{{\Large mK}}
\psfrag{Fiducial}[c][c][1][0]{{\Large {\bf {\textcolor{white}{Fiducial}}}}}
\psfrag{Clumping}[c][c][1][0]{{\Large {\bf {\textcolor{white}{Clumping}}}}}
\psfrag{UUVB Dom}[c][c][1][0]{{\Large {\bf {$\,\,\,\,\,\,\,\,\,\,\,\,\,\,$\textcolor{white}{UIB Dom}}}}}
\psfrag{SXR Dom}[c][c][1][0]{{\Large {\bf {\textcolor{white}{SXR Dom}}}}}
\psfrag{UV+SXR+UUVB}[c][c][1][0]{{\Large {\bf {$\,\,\,\,\,\,\,\,\,\,\,\,\,\,\,\,\,\,$\textcolor{white}{UV+SXR+UIB}}}}}
\psfrag{PL=2.0}[c][c][1][0]{{\Large {\bf {\textcolor{white}{PL 2.0}}}}}
\includegraphics[width=1.0\textwidth,angle=0]{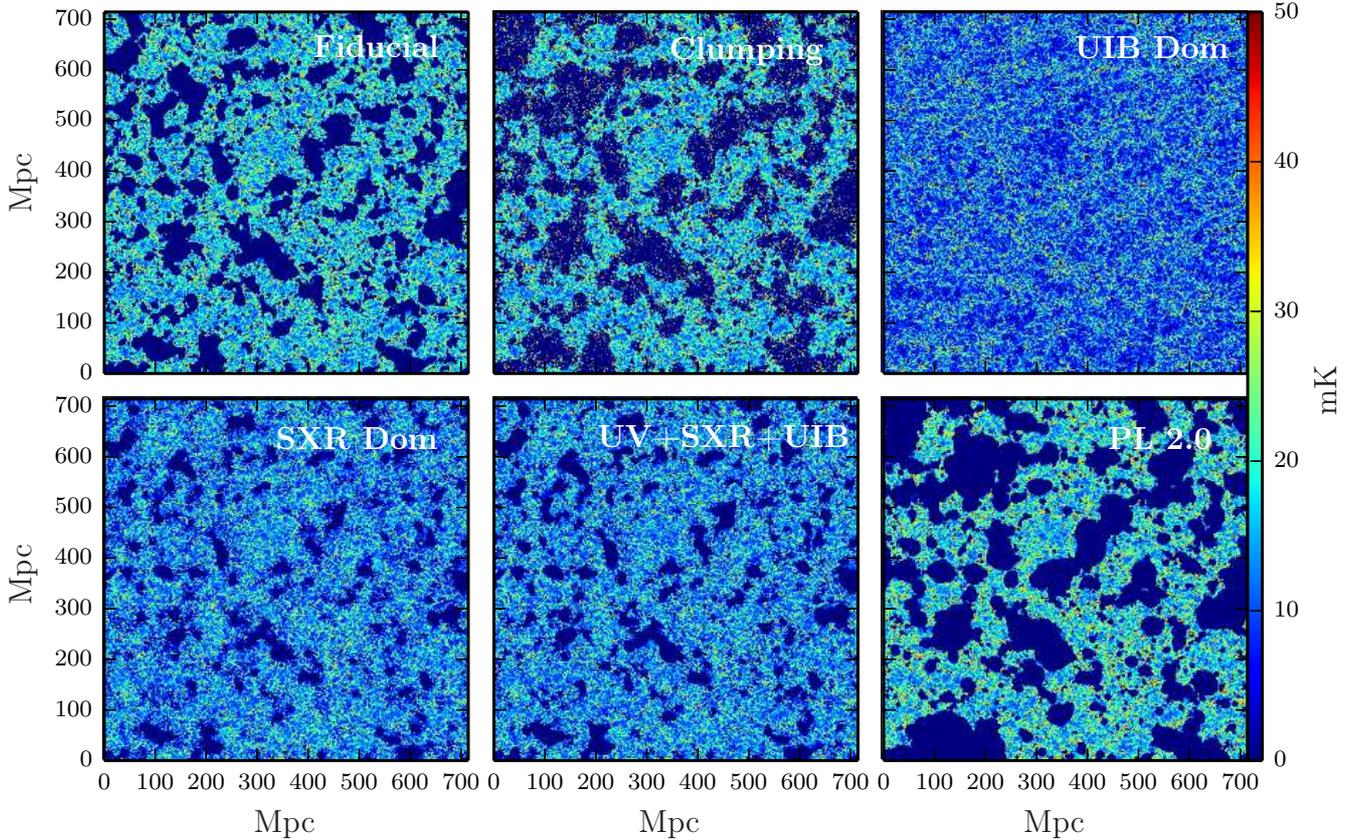}

\caption{Redshift space brightness temperature maps for six of our
  reionization scenarios. The line of sight is along the y-axis. The
  mass averaged neutral fraction for all the panels is $\xb = 0.5$.}
\label{fig:dt_maps}
\end{figure*}

\subsection{Generating redshift-space brightness temperature maps}
\label{sec:rsd_lc_map}
Following the steps described in Section \ref{sec:nbody} and
\ref{sec:reion} we have simulated coeval volumes of the matter
density, ionization and velocity fields at various redshifts for the
different models of reionization described in Table
\ref{tab:reion}. The density and ionization fields are then combined
to generate the brightness temperature maps assuming that {$T_s \gg
  T_{\mathrm{CMB}}$\footnote{Note that we have not taken into account
    the effect of fluctuations in the spin temperature. Spin
    temperature fluctuations due to Lyman-$\alpha$ pumping and heating
    by X-ray sources can affect the 21-cm brightness temperature
    fluctuations significantly during the early stages of the EoR
    \citep{mao12,mesinger13,ghara14}. However, as we discuss in the
    later parts of this paper (Section \ref{sec:spin_temp}), the
    anisotropy in the 21-cm power spectrum will not be strongly
    affected by this, especially once the early phase of EoR is over
    (i.e. when $\xb \leq 0.95$).}, where $T_s$ and $T_{\mathrm{CMB}}$
  are the spin temperature and the CMB temperatures respectively. We
  then take into account the peculiar velocities in order to construct
  the redshift-space signal. By redshift space, we mean the space that
  will be reconstructed by an observer assuming that all redshifts are
  purely due to the Hubble expansion. An emitter with a line-of-sight
  peculiar velocity $v_{\parallel}$ at a real-space position
  $\mathbf{r}$ will be translated to an apparent, redshift-space
  position $\mathbf{s}$ following
\begin{equation}
  \label{eq:redshift_translation}
  {\mathbf s} = {\mathbf r} + \frac{1 + z}{H(z)}v_{\parallel} \hat{r}
\end{equation}
For a more in-depth explanation of 21-cm redshift-space distortions,
see e.g.\ \citet{mao12}, \citet{jensen13} or \citet{majumdar13}.

We implement the redshift-space distortions using the same method as
in \cite{jensen13}. This method splits each cell into $n$ smaller
sub-cells along the line-of-sight and assigns each sub-cell the
brightness temperature $\delta T_{\mathrm{b}}({\mathbf r})/n$, where
$\delta T_{\mathrm{b}}({\mathbf r})$ is the temperature of the host
cell. It then moves the sub-cells according to Equation
\eqref{eq:redshift_translation}, and regrids them to the original
resolution. Here, we use $n=50$, which gives results that are accurate
down to approximately one-fourth of the Nyquist wave number, or $k
\lesssim k_{\mathrm{N}}/4 = \upi/4 \times 600/(714 \; \rm{Mpc}) = 0.66
\; \rm{Mpc}^{-1}$ \citep{mao12,jensen13}. Figure \ref{fig:dt_maps}
shows the resulting redshift-space brightness temperature maps for six
of the seven different reionization scenarios considered here (at a
time when the reionization had reached 50\%).

\section{Quantifying the power spectrum anisotropy}
\label{sec:rsd}

Since only the line-of-sight component of the peculiar velocities
affects the 21-cm signal, redshift-space distortions will make the
observed 21-cm power spectrum anisotropic. To quantify this
anisotropy, it is common to introduce a parameter $\mu$, defined as
the cosine of the angle between a wave number $\mathbf{k}$ and the
line-of-sight. The redshift-space power spectrum of the EoR 21-cm
signal at large scales can be written as a fourth-order polynomial in
$\mu$ under the quasi-linear approximation for the signal presented by
\citet{mao12}. This quasi-linear model is an improvement on the linear
model of \citet{bharadwaj05} and \citet{barkana05} for the 21-cm
signal. In the quasi-linear model one assumes the density and velocity
fluctuations to be linear in nature but the fluctuations in the
neutral fraction are considered to be non-linear. The expression for
the 21-cm power spectrum, $P^{\mathrm{s}} (k,\mu)$, becomes:
\begin{align}
  \label{eq:rsd_qlin}
  \nonumber
  P^{\mathrm{s}}(k, \mu) = \overline{\delta T_b}^2(z) &\left[ P_{\rhohi,\rhohi}(k)\right. +\\
  &\left.+ 2\mu^2 P_{\rhohi, \rhoh}(k) + \mu^4
    P_{\rhoh,\rhoh}(k) \right],
\end{align}
where $\rhohi$ is the neutral hydrogen density, and $\rhoh$ is
the total hydrogen density.

It has been suggested that this form of de-composition can be used to
extract the cosmology from the astrophysics on large spatial scales,
by extracting the coeffiecient of the $\mu^4$-term in Equation
\eqref{eq:rsd_qlin} \citep{barkana05,mcquinn06,shapiro13}. However,
this is very challenging due to the high level of cosmic variance at
these length scales and also due to the fact that this decomposition
is in a non-orthonormal basis.

However, the astrophysical information contained in the $\mu^2$ term
is also interesting. This term is determined by the cross-power
spectrum of the neutral matter density and the total matter density,
which is a measure of the ``inside-outness'' of the reionization
topology \citep{majumdar13,jensen13,ghara14}. However, extracting this
term from noisy data using Equation \eqref{eq:rsd_qlin} is also
challenging since power tends to leak over between the $\mu^2$ and
$\mu^4$ terms \citep{jensen13}.

One way to deal with this is to measure the sum of the $\mu^2$ and
$\mu^4$ terms \citep{jensen13}. This quantity is much more resistant
to noise. It will contain a mixture of cosmological and astrophysical
information but since the cosmological matter power spectrum evolves only
slowly and monotonically, the \emph{evolution} of the sum of the $\mu$
terms will be mostly determined by the $\mu^2$ term.

A different approach is to instead expand the power spectrum in the
orthonormal basis of Legendre polynomials---a well known
approach in the field of galaxy redshift surveys
\citep{hamilton92,hamilton98,cole95}. In this representation, the
power spectrum can be expressed as a sum of the even multipoles of
Legendre polynomials:
\begin{equation}
    P^{\mathrm{s}}(k,\mu) = \sum_{l\;\mathrm{even}} \Pl(\mu) P_l^{\mathrm{s}}(k).
  \label{eq:legendre}
\end{equation}
From an observed or simulated 21-cm power spectrum,
one can calculate the angular multipoles $P_l^{\mathrm{s}}$:
\begin{equation}
    P_l^{\mathrm{s}}(k) = \frac{2l + 1}{4\pi} \int \Pl(\mu) P^{\mathrm{s}}(k) \ud \Omega .
  \label{eq:multipole}
\end{equation}
The integral is done over the entire solid angle to take into account
all possible orientations of the ${\bf k}$ vector with the
line-of-sight direction. The estimation of each multipole moment
through Equation \eqref{eq:multipole} will be independent of the
other, as this representation is in an orthonormal basis.

Under the quasi-linear model of \citet{mao12}, only the first three
even multipole moments will have non-zero values
\citep{majumdar13,majumdar14}:
\begin{align}
    P^{\mathrm{s}}_0 &= \overline{\delta T_b}^2 \left[ \frac{1}{5} P_{\rhoh,\rhoh} + P_{\rhohi,\rhohi} + \frac{2}{3}P_{\rhohi,\rhoh} \right] \label{eq:legendre_P0} \\
    P^{\mathrm{s}}_2 &= 4\overline{\delta T_b}^2 \left[ \frac{1}{7} P_{\rhoh,\rhoh} + \frac{1}{3}P_{\rhohi,\rhoh} \right] \label{eq:legendre_P2} \\
    P^{\mathrm{s}}_4 &= \frac{8}{35}\overline{\delta T_b}^2
  P_{\rhoh,\rhoh} \label{eq:legendre_P4}
\end{align}
The monopole moment $P^{\mathrm{s}}_0$ is, by definition (Equation
\ref{eq:multipole}), the spherically averaged power spectrum. We
see that the quadrupole moment, $\Ps$, is a linear combination of
$P_{\rhoh,\rhoh}$ and $P_{\rhohi,\rhoh}$, just as the sum of the
$\mu^2$ and $\mu^4$ terms in Equation \eqref{eq:rsd_qlin}. Both the
sum of the $\mu^2$ and $\mu^4$ terms in Equation \eqref{eq:rsd_qlin}
and $P^{\mathrm{s}}_2$ in Equation \eqref{eq:legendre_P2} thus contain the same
physical information. However, in contrast to the decomposition of the
power spectrum in terms of powers of $\mu$, the expansion in the
Legendre polynomials is in an orthonormal basis, which means that the
uncertainty in the estimates of one moment will be uncorrelated with
the uncertainty in the estimates of the other moments. Therefore, 
the multipole moments should be easier to extract from noisy data. This
also implies that, if measured with statistical significance, each
multipole moment can be used as an independent and complementary
estimator of the cosmological 21-cm signal.

\begin{figure}
  \includegraphics[width=.5\textwidth,angle=0]{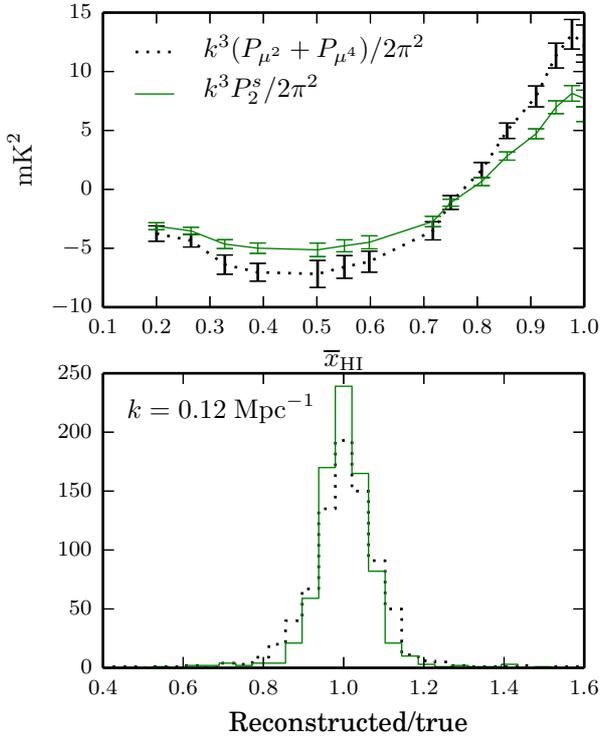}
  \caption{Comparison of the two different ways of measuring the anisotropy. 
      The top panel shows the anisotropy reconstructed
    from noisy data (see the text for details) as a function of
    neutral fraction. The error bars show the $2\sigma$ variation for
    50 different noise realizations. The bottom panel shows a
    histogram of the reconstructed noisy values divided by the true
    values. A narrower distribution around $1$ is an indication of a
    better reconstruction.}
  \label{fig:anisotropy_methods}
\end{figure}
\begin{figure*}
  \includegraphics[width=1.0\textwidth,angle=0]{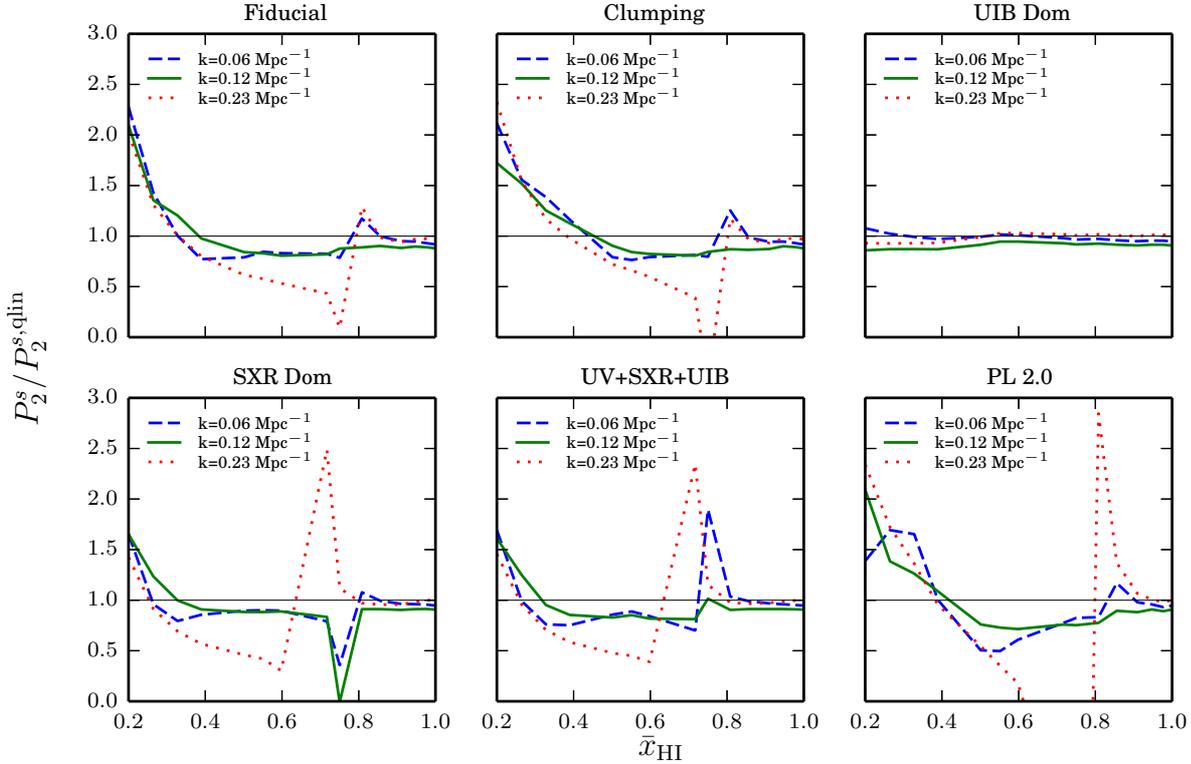}
  \caption{The ratio between the quadrupole moment measured from the
    simulated 21-cm signal, and the quasi-linear approximation,
    calculated from Equation \eqref{eq:legendre_P2} for our different
    reionization scenarios.}
  \label{fig:qlin_ratio}
\end{figure*}
To test this, we took simulated brightness temperature volumes from
the fiducial model (see Section \ref{sec:reion}) and added 50
different realizations of Gaussian noise ({\bf later on, we will make
predictions with realistic noise for LOFAR and the SKA}). After calculating the power
spectra of the noisy data volumes, we fit a fourth-order polynomial in
$\mu$ using standard least-squares fitting, and also estimated the
$\Ps$ moment using Equation \eqref{eq:multipole}. The results of this
test are shown in Figure \ref{fig:anisotropy_methods}. The top panel
shows the measured values of the anisotropy using the two methods. The
solid lines show the true, noise-free values, and the error bars show
the $2\sigma$ variation in the noisy measurements across the different
noise realizations. In the bottom panel, we show a histogram of the
ratio between the measured and the true value for two methods. While
the difference is not dramatic, we see here that the $\Ps$ is slightly
more resistant to noise than the sum of the $\mu$ terms. Because of
this and also due to the fact that each multipole moment represented
in the orthonormal basis of of Legendre polynomials will be
independent of the other, we will use the quadrupole moment $\Ps$, as
our measure of power spectrum anisotropy for the rest of this paper.

Fluctuations in spin temperature $T_s$, introduced due to the heating
by the very early astrophysical sources will affect the monopole
moment ($P^{\mathrm{s}}_0$) or the spherically averaged power spectrum
significantly \citep{fialkov14a,ghara14}. However we expect that they
will not have much impact on the anisotropy term or the quadrupole
moment ($\Ps$). We direct the reader to Section \ref{sec:spin_temp}
for further details.

\subsection{How good is the quasi-linear model for
  interpreting the anisotropy?}
\label{sec:qlin}
The quadrupole moment gives us a measure of the power spectrum
anisotropy. Ideally, we would want to connect this measure to the topology of
reionization. The quasi-linear approximation gives us a way to do this
in the early stages of reionization and for large length scales. When
this approximation holds, Equation \eqref{eq:legendre_P2} tells us
that $\Ps$ is a weighted sum of the matter power spectrum and the
cross-power spectrum of the total and neutral matter densities. This
cross-power spectrum measures how strongly correlated the \HI and
matter distributions are. If reionization is strongly inside-out, this
quantity will decrease quickly as reionization progresses, and take on
a negative value (see e.g.\ \citealt{jensen13,majumdar13,ghara14}).

To some degree, the success of the quasi-linear approximation depends
on the source model. In Figure\ \ref{fig:qlin_ratio}, we show the
ratio between the true value of $\Ps$ and the quasi-linear
expectation, constructed by first calculating $P_{\rhoh,\rhoh}$ and
$P_{\rhohi,\rhoh}$ and then combining them according to Equation\
\eqref{eq:legendre_P2}. We see that at $k=0.12$ Mpc$^{-1}$, the
approximation works rather well (i.e.\ the ratio is close to 1) for
all scenarios up to around $\xb \sim 0.5$. For larger spatial scales,
the effect of sample variance becomes significant, while at smaller
spatial scales, non-linearities start influencing the results. The
quasi-linear approximation for the signal works poorly for the PL 2.0
scenario, but works almost perfectly until very late stages of EoR for
the UIB Dominated scenario. Around $\xb=0.7$, the quadrupole moment
crosses zero, giving rise to sudden jumps in the ratio.

Note that the quadrupole moment of the power spectrum is a quantity
that can always be calculated from the data, even when the signal is
highly non-linear and also regardless of the assumed model for the
signal. The quasi-linear model is simply a tool to help with
interpreting the quantity. The quadrupole moment will still be useful
as an independent estimator of the signal and as an observable to
distinguish between different source models even in the non-linear
regime. Here, all we can do is look at how this estimator evolves in
simulations with different reionization scenarios, which we discuss in
the next section.

\begin{figure}
\psfrag{xh1}[c][c][1][0]{{\LARGE $\xb$}}
\psfrag{P0}[c][c][1][0]{{\LARGE $k^3P^s_{0}/(2\pi^2)\, ({\rm mK^2})$}}
\psfrag{k = 0.122 Mpc-1}[c][c][1][0]{{\Large $k=0.12\,{\rm Mpc}^{-1}$}}
\psfrag{fidu}[c][c][1][0]{Fiducial$\,\,\,\,\,\,\,\,$}
\psfrag{fidu inh}[c][c][1][0]{{Clumping$\,\,\,\,$}}
\psfrag{xray=0.8}[c][c][1][0]{{UIB Dom$\,$}}
\psfrag{xrays=0.8}[c][c][1][0]{{SXR Dom}}
\psfrag{xraysh=0.5}[c][c][1][0]{{UV+SXR+UIB$\,\,\,\,\,\,\,\,\,\,\,$}}
\psfrag{pl=2.0}[c][c][1][0]{{PL 2.0}}
\psfrag{pl=3.0}[c][c][1][0]{{PL 3.0}}
\includegraphics[width=.5\textwidth,angle=0]{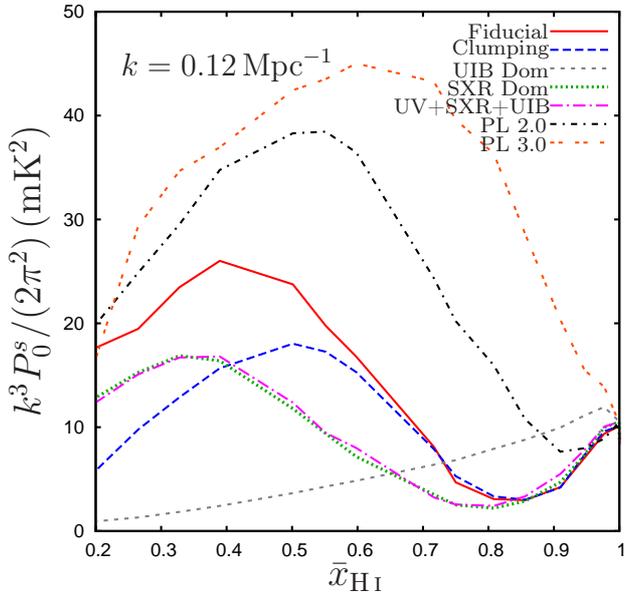}
\caption{The monopole moment of the power spectrum for all of our
  reionization scenarios as a function of the global neutral fraction
  at $k=0.12$ Mpc$^{-1}$.}
\label{fig:p0}
\end{figure}
\section{Redshift-space distortions in different reionization scenarios}
\label{sec:result}
In Figure \ref{fig:p0}, we show the monopole moment of the power
spectrum, $P^{\mathrm{s}}_0$---which is by definition the spherically
averaged power spectrum---for the reionization scenarios listed in
Table \ref{tab:reion}. From here onwards we show all our results for
the wave number $k=0.12$Mpc$^{-1}$. It is expected that LOFAR will be
most sensitive around this length scale \citep{jensen13} and also the
quasi-linear approximations seems to work quite well for this length
scale (Figure \ref{fig:qlin_ratio}). The quantity $P^{\mathrm{s}}_0$
has been studied extensively before (e.g.\
\citealt{santos10,mesinger11,mao12,majumdar13,jensen13,ghara14}
etc.), and the trends we are observing here mostly agree with these
previous studies.

At the large scales we consider here, the behaviour of
$P^{\mathrm{s}}_0$ can be readily understood by the quasi-linear
approximation (Equation \ref{eq:legendre_P0}). Initially, the
brightness temperature fluctuations mostly come from fluctuations in
the \HI distribution, and $P_0^{\mathrm{s}}$ decreases as the density
peaks are ionized. Later on, at a neutral fraction around $0.3-0.5$
(depending on the reionization scenario), a peak in $P_0^{\mathrm{s}}$
appears, as large ionized regions form and become the main
contributors to the brightness temperature fluctuations. This peak is
especially pronounced in the PL scenarios, where the ionized bubbles
are particularly large. However, it is completely absent in the UIB
Dominated scenario, since due to the presence of a uniform ionizing
background the \HI distribution at all stages follows the product of
the matter distribution and the mean brightness temperature, which
causes $P_0^{\mathrm{s}}$ to fall off monotonically with decreasing
$\xb$.

\begin{figure}
\psfrag{xh1}[c][c][1][0]{{\LARGE $\xb$}}
\psfrag{P2}[c][c][1][0]{{\LARGE $k^3P^s_{2}/(2\pi^2)\, ({\rm mK^2})$}}
\psfrag{k = 0.122 Mpc-1}[c][c][1][0]{{\Large $k=0.12\,{\rm Mpc}^{-1}$}}
\psfrag{fidu}[c][c][1][0]{Fiducial$\,\,\,\,\,\,\,\,$}
\psfrag{fidu inh}[c][c][1][0]{{Clumping$\,\,\,\,$}}
\psfrag{xray=0.8}[c][c][1][0]{{UIB Dom$\,$}}
\psfrag{xrays=0.8}[c][c][1][0]{{SXR Dom}}
\psfrag{xraysh=0.5}[c][c][1][0]{{UV+SXR+UIB$\,\,\,\,\,\,\,\,\,\,\,$}}
\psfrag{pl=2.0}[c][c][1][0]{{PL 2.0}}
\psfrag{pl=3.0}[c][c][1][0]{{PL 3.0}}
\includegraphics[width=.5\textwidth,angle=0]{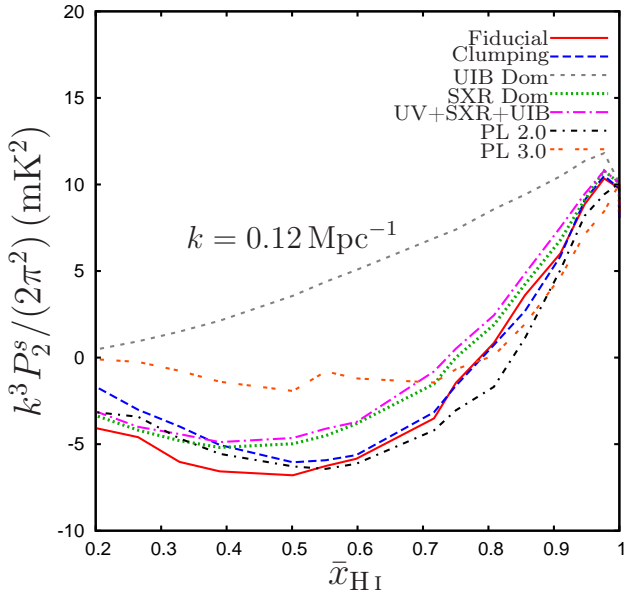}
\caption{The quadrupole moment of the power spectrum for all of our
  reionization scenarios as a function of the global neutral fraction
  at $k=0.12$ Mpc$^{-1}$.}
\label{fig:p2}
\end{figure}
\begin{figure}
\psfrag{xh1}[c][c][1][0]{{\LARGE $\xb$}}
\psfrag{Pcross}[c][c][1][0]{{\LARGE $\overline{\delta T_{\mathrm{b}}}^2k^3P_{\rhoh, \rho_{\HI}}/(2\pi^2)\, ({\rm mK^2})$}}
\psfrag{k = 0.122 Mpc-1}[c][c][1][0]{{\Large $k=0.12\,{\rm Mpc}^{-1}$}}
\psfrag{fidu}[c][c][1][0]{Fiducial$\,\,\,\,\,\,\,\,$}
\psfrag{fidu inh}[c][c][1][0]{{Clumping$\,\,\,\,$}}
\psfrag{xray=0.8}[c][c][1][0]{{UIB Dom$\,$}}
\psfrag{xrays=0.8}[c][c][1][0]{{SXR Dom}}
\psfrag{xraysh=0.5}[c][c][1][0]{{UV+SXR+UIB$\,\,\,\,\,\,\,\,\,\,\,$}}
\psfrag{pl=2.0}[c][c][1][0]{{PL 2.0}}
\psfrag{pl=3.0}[c][c][1][0]{{PL 3.0}}
\psfrag{den}[c][c][1][0]{{{\tiny $\overline{\delta T_{\mathrm{b}}}^2k^3P_{\rhoh, \rhoh}/(2\pi^2)\,\,\,\,\,\,\,\,\,\,\,\,\,\,\,\,\,\,\,\,\,\,\,\,\,\,\,\,\,\,\,\,\,\,\,\,\,\,\,\,\,\,\,$}}}
\includegraphics[width=.5\textwidth,angle=0]{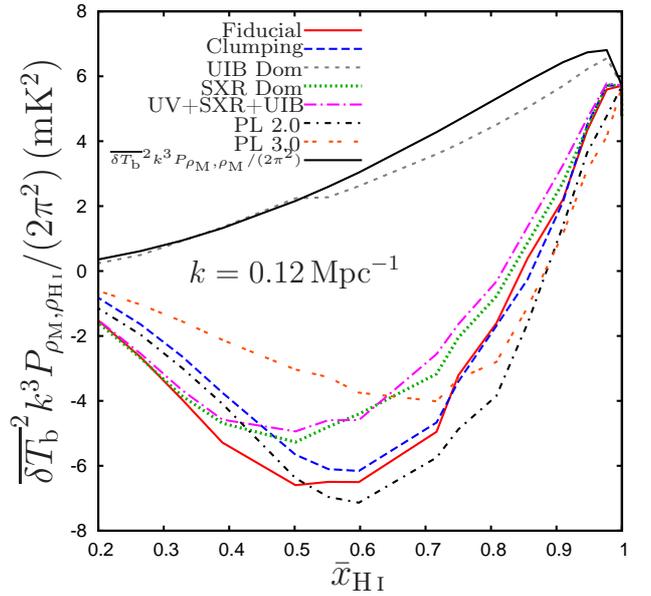}
\caption{The matter and \HI density cross-power spectrum as a function
  of the global neutral fraction for all the reionization scenarios at
  $k=0.12$ Mpc$^{-1}$. The black solid line shows the evolution of the
  matter power spectrum with $\xb$, which is same for all the
  reionization scenarios considered here.}
\label{fig:p_cross}
\end{figure}

Figure \ref{fig:p2} shows the quadrupole moment, $\Ps$, which we use
as a measure of the power spectrum anisotropy. It was also studied in
\citet{majumdar13} and \citet{majumdar14} for EoR scenarios similar to
the fiducial and clumping scenarios described in this paper. As we saw
previously (Section \ref{sec:qlin}), the behaviour of $\Ps$ can be
understood to some degree using the quasi-linear model (Equation
\ref{eq:legendre_P2}), where $\Ps$ is a linear combination of the
matter power spectrum $P_{\rhoh,\rhoh}$ and the cross-power spectrum
between the total matter density and the neutral matter density,
$P_{\rhohi,\rhoh}$. We show the product of the square of the mean
brightness temperature ($\overline{\delta T_{\mathrm{b}}}^2$) and the
cross-power spectrum for the different reionization scenarios in
Figure \ref{fig:p_cross}, along with the product of the
$\overline{\delta T_{\mathrm{b}}}^2$ and the matter power spectrum
(which is the same for all the scenarios). Since the matter power
spectrum increases slowly and monotonously and $\overline{\delta
  T_{\mathrm{b}}}^2$ decreases rather rapidly with the decreasing
$\xb$, the evolution of the term $\overline{\delta T_{\mathrm{b}}}^2
P_{\rhoh, \rhoh}$ is driven by the evolution of $\overline{\delta
  T_{\mathrm{b}}}^2$. As the contribution from the term
$\overline{\delta T_{\mathrm{b}}}^2 P_{\rhoh, \rhoh}$ is the same for
all the reionization scenarios considered here, the major differences
between $\Ps$ for different reionization scenarios come from the
differences in the evolution of $\overline{\delta T_{\mathrm{b}}}^2
P_{\rhoh, \rho_{\HI}}$.
Going back to Figure \ref{fig:p2}, for most scenarios $\Ps$ initially
increases in strength and then starts falling.  This is because when
the IGM is completely neutral, $\rhohi = \rhoh$ and $P_{\rhohi,\rhoh}$
will increase when the matter fluctuations grow. As the most massive
peaks are ionized, however, $\rhohi$ becomes anti-correlated with
$\rhoh$ and $\Ps$ becomes negative. The stronger the
anti-correlation---i.e.\, the more inside-out the reionization
topology is---the more negative $\Ps$ becomes. The scenarios that
stand out from the rest are again the UIB Dominated scenario, where
the reionization is more outside-in than inside-out, and the PL 3
scenario. In the latter, reionization is driven mainly by a few
massive sources that produce very large regions of ionized hydrogen
(\HII). Since these \HII bubbles are so large, the correlation between
the total matter and \HI field becomes very weak at large scales, and
so $\Ps$ remains close to zero for most of the reionization history.

Since $P_0^{\mathrm{s}}$ and $\Ps$ are two independent but
complementary measurements of the 21-cm signal, one can thus visualize
the evolution of the 21-cm signal in a reionization scenario as a
trajectory in the phase space of $P_0^{\mathrm{s}}$ and $\Ps$.  In
Figure \ref{fig:p2p0_vec} we show the reionization scenarios
considered here in such a phase space diagram. Looking first at the
fiducial scenario, it initially moves upwards to the right, with both
$P_0^{\mathrm{s}}$ and $\Ps$ increasing as the matter fluctuations
grow. At a ionization fraction of a few percent, it starts moving
downward in an arc, eventually reaching another turn-around point at
$\xb \sim 0.4$. The point where the \HI fluctuations are at their
strongest seems to roughly coincide with the point where $\Ps$ reaches
its minimum. After this, the trajectory approaches the ($0,0$) point
of the phase space, as the IGM becomes more and more ionized and the
21-cm signal starts to disappear.

From Figures \ref{fig:p2p0_vec} and \ref{fig:p2}, we see that the
power spectrum anisotropy (i.e.\ $\Ps$) evolves remarkably similarly
in most of the reionization scenarios, even when the
spherically averaged power spectrum (i.e.\ $P_0^{\mathrm{s}}$) differs. The slope
is very similar in the early stages of reionization, and the minima in
$\Ps$ occur at around $\xb \sim 0.5$ for all reionization scenarios
except the extreme UIB Dominated scenario. 

In Appendix \ref{sec:appendix_b} we show that the strongest
contributing factor to the evolution of $\Ps$ is the phase
difference between the matter and the \HI fields. As long as a major
portion of the ionizing flux originates from the high density regions
(or the collapsed objects), the phase difference between the two
fields will evolve in roughly the same way regardless of the strength
of the \HI fluctuations. Only in extreme scenarios such as our
UIB Dominated scenario---where the flux is evenly distributed---does the
phase difference deviate from the other scenarios.

Thus, the power spectrum anisotropy will probably not be useful in
telling different source models apart, unless the sources are extreme
in some sense. However, it appears to offer a robust way of measuring
the history of reionization. We discuss this in more detail in the
next section.
\begin{figure}
\psfrag{P0}[c][c][1][0]{{\LARGE $k^3P^s_{0}/(2\pi^2)\, ({\rm mK^2})$}}
\psfrag{P2}[c][c][1][0]{{\LARGE $k^3P^s_{2}/(2\pi^2)\, ({\rm mK^2})$}}
\psfrag{k = 0.122 Mpc-1}[c][c][1][0]{{\Large $k=0.12\,{\rm Mpc}^{-1}$}}
\psfrag{fidu}[c][c][1][0]{Fiducial$\,\,\,\,\,\,\,\,$}
\psfrag{fidu inh}[c][c][1][0]{{Clumping$\,\,\,\,$}}
\psfrag{xray=0.8}[c][c][1][0]{{UIB Dom$\,$}}
\psfrag{xrays=0.8}[c][c][1][0]{{SXR Dom}}
\psfrag{xraysh=0.5}[c][c][1][0]{{UV+SXR+UIB$\,\,\,\,\,\,\,\,\,\,\,$}}
\psfrag{pl=2.0}[c][c][1][0]{{PL 2.0}}
\psfrag{pl=3.0}[c][c][1][0]{{PL 3.0}}
\psfrag{9.999960e-01}[c][c][1][0]{{\tiny 0.9999}}
\psfrag{9.995470e-01}[c][c][1][0]{{\tiny 0.9995}}
\psfrag{9.769730e-01}[c][c][1][0]{{\tiny 0.98}}
\psfrag{9.470890e-01}[c][c][1][0]{{\tiny 0.95}}
\psfrag{9.098670e-01}[c][c][1][0]{{\tiny 0.91}}
\psfrag{8.557260e-01}[c][c][1][0]{{\tiny 0.86}}
\psfrag{8.082530e-01}[c][c][1][0]{{\tiny 0.81}}
\psfrag{7.502640e-01}[c][c][1][0]{{\tiny 0.75}}
\psfrag{7.169550e-01}[c][c][1][0]{{\tiny 0.72}}
\psfrag{5.979810e-01}[c][c][1][0]{{\tiny 0.60}}
\psfrag{5.514900e-01}[c][c][1][0]{{\tiny 0.55}}
\psfrag{5.013450e-01}[c][c][1][0]{{\tiny 0.50}}
\psfrag{3.895350e-01}[c][c][1][0]{{\tiny 0.39}}
\psfrag{3.278210e-01}[c][c][1][0]{{\tiny 0.33}}
\psfrag{2.646340e-01}[c][c][1][0]{{\tiny 0.26}}
\psfrag{2.000000e-01}[c][c][1][0]{{\tiny 0.20}}
\includegraphics[width=.5\textwidth,angle=0]{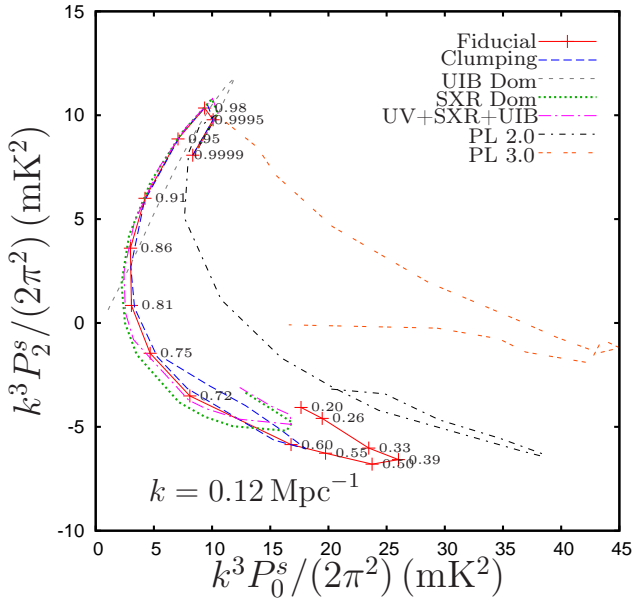}
\caption{Trajectories of the different reionization scenarios in
  $P^s_{2}-P^s_{0}$ phase space for $k=0.12$ Mpc$^{-1}$. The
  corresponding values of $\xb$ have been printed on the trajectory of
  the fiducial scenario show the global neutral fraction. For clarity,
  this has been shown only for the fiducial scenario.}
\label{fig:p2p0_vec}
\end{figure}

\subsection{The effects of spin temperature fluctuations}
\label{sec:spin_temp} 

In the reionization scenarios discussed here we have not considered
the effect of spin temperature fluctuations due to Lyman-$\alpha$
pumping and heating by X-ray sources. These effects can influence the
21-cm brightness temperature fluctuations significantly during the
early stages of EoR, if the heating of the IGM is late
\citep{fialkov14a}. This can also affect the line-of-sight anisotropy
in the signal \citep{ghara14,fialkov15,ghara15}. In a late-heating
scenario, if we define the spin temperature fluctuations in the \HI
distribution as
\begin{equation}
\eta(z,{\bf x}) = 1- \frac{T_{{\rm CMB}}(z)}{T_{{\rm S}}(z,{\bf x})},
\label{eq:eta}
\end{equation}
then the angular multipole moments of the redshift space power
spectrum under the quasi-linear approximations will take the form
\begin{align}
  P^s_0 &= \overline{\delta T_b }^2(z) \left[ \frac{1}{5} P_{\rhoh,
      \rhoh} + P_{\rhohi, \rhohi} + P_{\eta, \eta} + \right. \label{eq:eta_P0}\\
  \nonumber
  &\left. + 2 P_{\eta, \rhohi} + \frac{2}{3} P_{\rhohi, \rhoh} + \frac{2}{3} P_{\eta, \rhoh} \right] \\
  P^s_2 &= 4\, \overline{\delta T_b }^2(z) \left[ \frac{1}{7} P_{\rhoh, \rhoh}  + \frac{1}{3} P_{\rhohi, \rhoh} + \frac{1}{3} P_{\eta, \rhoh} \right] \label{eq:eta_P2}\\
  P^s_4 &= \frac{8}{35}\, \overline{\delta T_b }^2(z)\, P_{\rhoh,
    \rhoh} \label{eq:eta_P4}
\end{align}

As shown by \citet{fialkov14a,ghara14} and \citet{ghara15}, the spin
temperature fluctuations affect the monopole moment of the power
spectrum (i.e.\ the spherically averaged power spectrum) most
severely. Figures 8 and 9 in \citet{ghara14} show that the amplitude
of the power spectrum gets amplified significantly at all length
scales (by at least two orders of magnitude) during the cosmic dawn
(when the first X-ray sources emerge) and there is a prominent dip in
its amplitude at large length scales during the early phases of
reionization (i.e.\ when $\xb \approx 0.95$). This behaviour of the
spherically averaged power spectrum is due to the fact that during the
cosmic dawn and the early stages of the EoR, the auto power spectrum
of the spin temperature fluctuations, $P_{\eta, \eta}$, makes the
dominant contribution to the monopole moment (see figure 10 of
\citealt{ghara14}). This term will be zero when one assumes that $T_s
\gg T_{\mathrm{CMB}}$. They have not estimated any higher order
multipole moments of the power spectrum from their simulations.

However, as we can see from Equation \eqref{eq:eta_P2}, the $P_{\eta,
  \eta}$ term does not contribute to the quadrupole moment which we
use in this paper to quantify the redshift space anisotropy in the
power spectrum. The only contribution from spin temperature
fluctuations to $P^s_2$ is the cross-power spectrum of the spin
temperature fluctuations and the matter density fluctuations,
$P_{\eta, \rhoh}$. In the case of late heating, \citet{ghara14} find
that the $P_{\eta, \rhoh}$ term is comparable to $P_{\rhohi, \rhoh}$
at large scales during the very early stages of the EoR (i.e.\ $\xb
\sim 1.0 - 0.95$), but later on, it becomes few orders of magnitude
smaller. The evolution of $P_{\eta, \rhoh}$ with redshift and $\xb$ as
shown in \citet{ghara14}, may vary depending on the properties of the
heating sources. These results thus suggest that the quadrupole moment
is not completely immune to spin temperature fluctuations, but it is
much less sensitive than the monopole moment.

\section{Observability of the redshift-space anisotropy}
Having seen how the power spectrum anisotropy behaves under idealized
conditions, we now explore some of the complications one will
encounter when attempting to observe this effect.

\begin{figure*}
\psfrag{xh1}[c][c][1][0]{{\LARGE $\xb$}}
\psfrag{P2}[c][c][1][0]{{\Large $k^3P^s_{2}/(2\pi^2)\, ({\rm mK^2})$}}
\psfrag{P0}[c][c][1][0]{{\Large $k^3P^s_{0}/(2\pi^2)\, ({\rm mK^2})$}}
\psfrag{k = 0.122 Mpc-1}[c][c][1][0]{{$k=0.12\,{\rm Mpc}^{-1}$}}
\psfrag{Fiducial}[c][c][1][0]{Fiducial}
\psfrag{xray=0.8}[c][c][1][0]{{UIB Dom}}
\psfrag{pl=3.0}[c][c][1][0]{{PL 3.0}}
\psfrag{coeval}[c][c][1][0]{Coeval}
\psfrag{10 MHz}[c][c][1][0]{10 MHz}
\psfrag{15 MHz}[c][c][1][0]{15 MHz}
\psfrag{20 MHz}[c][c][1][0]{20 MHz}
\psfrag{25 MHz}[c][c][1][0]{25 MHz}
\includegraphics[width=1.\textwidth,angle=0]{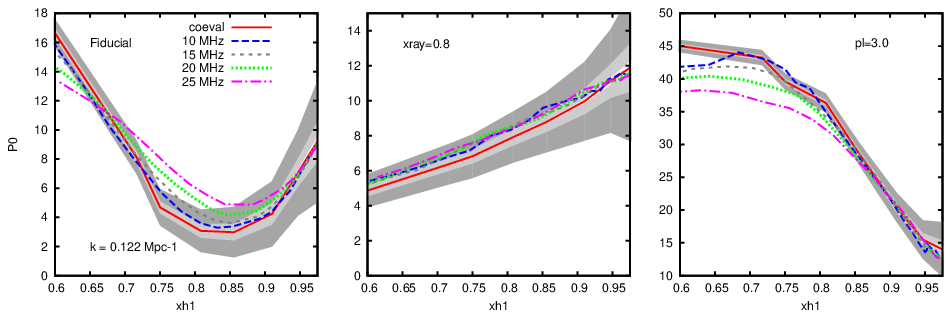}
\includegraphics[width=1.\textwidth,angle=0]{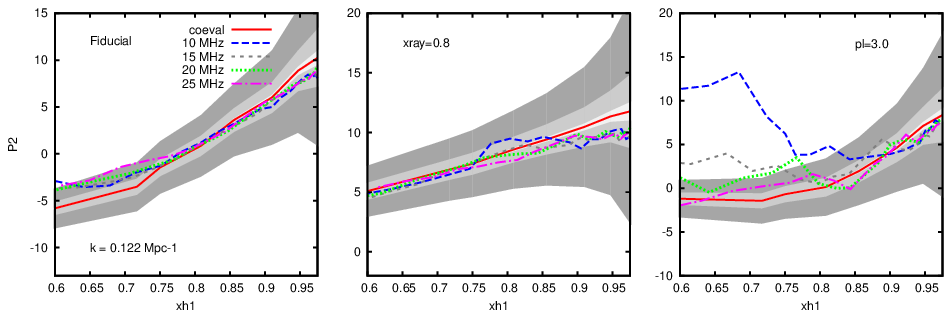}
\caption{The monopole and quadrupole moments of the power spectrum for
  three different reionization scenarios as a function of global
  neutral fraction at $k=0.12$ Mpc$^{-1}$ estimated from the light
  cone cubes with different frequency bandwidths. Note that for the
  light cone cube results the $\xb$ values quoted in the x-axis of
  these plots are the effective $\xb$ values within the sub-volume
  limited by the bandwidth. The shaded regions in three different
  shades of grey, dark to light, show the $1\sigma$ uncertainty in
  these estimates due to the system noise, measured in a bandwidth of
  $20$ MHz, after $1000$ and $3000$ hours of LOFAR and $100$ hours SKA
  observations, respectively.}
\label{fig:p2_lc}
\end{figure*}

\label{sec:obs}
\subsection{The lightcone effect}
In the previous section, we showed the 21-cm power spectra
estimated from so-called ``coeval'' data volumes, i.e.\ the direct
outputs from simulations, where the signal is at a fixed evolutionary
stage in the entire volume. Because of the finite travel time of
light, an observer will not be able to observe the signal like this.
Instead, the signal will be seen at a different evolutionary
stage at each observed frequency. This is known as the lightcone
effect.

Because of the lightcone effect, a three-dimensional measurement of
the 21-cm signal can never be done at a fixed single neutral
fraction. To measure, say, the power spectrum from a real observation,
one has to average over some frequency range (corresponding to a range
in neutral fraction). This averaging has been shown to have a small
effect on the spherically averaged power spectrum
\citep{datta12}{\footnote{The lightcone averaging could have a
    significant impact on the spherically averaged power spectrum,
    especially during the early stages of EoR, if one considers the
    spin temperature fluctuations in the signal due to the
    inhomogeneous heating by the X-ray sources and considers averaging
    the signal for a significantly large frequency bandwidth
    \citep{ghara15}.}}. It has also been shown that the lightcone
effect does not by itself introduce any significant additional
anisotropy in the 21-cm signal at measurable scales
\citep{datta14}. However, the averaging will have some effect on the
measurements of the anisotropy that is already present due to the
redshift-space distortions.

In Figure \ref{fig:p2_lc}, we show the monopole and the quadrupole
moments of the power spectrum estimated from lightcone volumes
generated as described in Section \ref{sec:rsd_lc_map}. We compare the
results for coeval data volumes with bandwidths of $10,\, 15,\, 20\,
{\rm and}\, 25$ MHz for the fiducial reionization scenario, along with
the two most extreme scenarios: the UIB Dominated and PL 3
scenarios. In general, a narrower bandwidth means that the effective
averaging is done over a smaller range of neutral fractions, and the
value of $\Ps$ and $P_0^{\mathrm{s}}$ will follow the coeval values more
closely. However, a narrow bandwidth will increase the sample
variance. 

Figure \ref{fig:p2_lc} shows that for the fiducial scenario, the
lightcone effect changes $\Ps$ only marginally, and only at low neutral
fractions. The effect is strongest for the other two scenarios, but 
this is mostly due to sample
variance, as the signal is being measured over a narrow slice of the
full volume. The lightcone effect on $P_0^{\mathrm{s}}$ is largest around $\xb
\sim 0.8$ for the fiducial model. For the other two reionization
scenarios it is more prominent at low neutral fractions. 

\subsection{Detector noise}
Any radio interferometric measurement of the 21-cm signal from the EoR
will face several obstacles, including detector noise, galactic and
extragalactic foreground sources and ionospheric disturbances. While a
full treatment of all these effects is beyond the scope of this paper,
we do investigate the fundamental limitations from detector noise and
foregrounds for LOFAR and the SKA.

To calculate the detector noise, we use the same method as in
\cite{jensen13}.  We begin by calculating the $u,v$ coverage for the
antenna distribution in the LOFAR core \citep{yatawatta13}. For the
SKA, we assume a Gaussian distribution of antennas within a radius of
around 2000 meters \citep{dewdney13}, assuming 150 antenna stations.

We fill the $u,v$ planes with randomly generated Gaussian noise with a
magnitude calculated using the formalism from \cite{mcquinn06}. We use
the same parameters for LOFAR as in \cite{jensen13}. For the SKA we
assume a total collecting area of $5\times 10^5$ m$^2$. This
collecting area, along with the assumptions for the baseline
distribution, is consistent with the current plans for the first-phase
version of the low-frequency part of the SKA (SKA1-LOW) \citep{koopmans15}.

Finally, we Fourier transform the noise in the $u,v$ plane to get the
noise in the image plane. We do this for many different frequencies to
get a noise lightcone with the same dimensions as our signal
lightcones. Here, we do not take into account the frequency dependence
of the $u,v$ coverage.

Using these simulated noise lightcones, we can then calculate the
power spectrum error due to detector noise.  In Figure
\ref{fig:p2_lc}, we show these errors for 1000 and 3000 hours of LOFAR
observations and 100 hours of SKA observations as shaded regions (in
three different shades of grey, dark to light, respectively). We see
that for LOFAR, the noise error for $\Ps$ is much higher than the
uncertainty due to the lightcone effect. Clearly, LOFAR will
require a long integration time to measure the evolution in
$\Ps$. For the SKA, however, the noise error is almost negligible,
even for relatively short integration times. In this case, the
lightcone effect also becomes important, at least at the later stages
of reionization.

\subsection{Foregrounds and PSF effects}

\begin{figure}
  \includegraphics[width=0.5\textwidth]{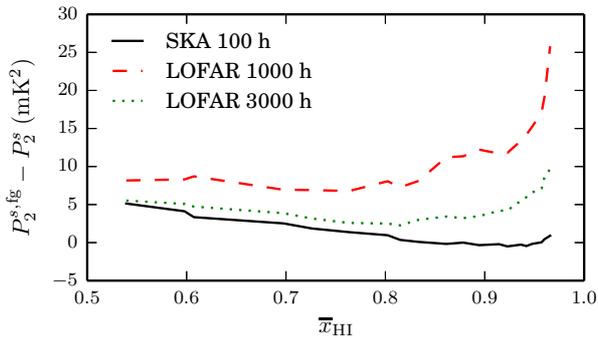}
  \caption{Effects of foreground subtraction on $\Ps$. The figure
    shows the difference between $\Ps$ measured after and before
    adding and subtracting foreground for a few different integration
    times for LOFAR and the SKA. The results are shown at $k=0.12$
    Mpc$^{-1}$ for the fiducial scenario, with the measurements
    carried out at 15 MHz bandwidth.}
  \label{fig:fg_effects}
\end{figure}

In addition to detector noise, there are many other issues that will
complicate the extraction of $\Ps$ from a real measurement. One
major obstacle is foreground emission from Galactic and extra-Galactic
sources. To properly measure the power spectrum anisotropy, the
foregrounds need to be removed from the data using some algorithm.

We simulate foreground emission from diffuse and localized Galactic
synchrotron radiation, Galactic free-free radiation and extragalactic
bright radio sources using the methods described in
\cite{jelic08,jelic10}. After adding these sources to the simulated
signal, we use the GMCA algorithm to model and remove them (see
\citealt{chapman2013} for a detailed description). GMCA (Generalized
Morphological Component Analysis) works by attempting to express the
full signal+foregrounds in a wavelet basis where only a few components
are non-zero. Since the signal is very weak compared to the
foregrounds, it will tend not to be included in the wavelet basis and
can thus be separated from the foregrounds.

A second complication is the beam, or the point-spread function (PSF),
of the instrument. The effect of the PSF is essentially a smoothing in
the sky plane, de-emphasizing large $k_{\perp}$ modes and introducing
an anisotropy in the signal. To properly study the feasibility of
extracting $\Ps$ from real measurements, we would need to model the
frequency-dependent PSF and compensate for its effects, which is
beyond the scope of this paper. Instead, we simply calculate the
change in $\Ps(\xb)$ after adding and subtracting foregrounds. This
change is shown in Figure \ref{fig:fg_effects} for a few different
integration times.  Here, $P_2^{\mathrm{s},\mathrm{fg}}$ denotes the
quadrupole moment after adding and subtracting foregrounds, while
$\Ps$ is the quadrupole with no foreground effects included. We show
only the absolute error since the relative error fluctuates wildly
where the signal crosses zero.

We see that for low noise levels, the foregrounds change the measured
$\Ps$ by a few mK$^2$. As seen in Figure \ref{fig:p2}, the signal
itself ranges between approximately $-5$ and $10$ mK$^2$.  For the
SKA, the foreground removal is almost perfect at high values of $\xb$,
while LOFAR observations appear to require long integration times to
properly measure $\Ps$. However, the GMCA algorithm used here is not
the only method for removing foregrounds, and it remains possible that
other algorithms perform better for this particular task.

\subsection{Reconstructing the reionization history}
As we saw in Figure \ref{fig:p2}, the quadrupole moment evolves very
predictably as a function of $\xb$; much more so than the monopole
moment or the spherically averaged power spectrum (Figure
\ref{fig:p0}). The only scenarios that deviate from the rest are the
UIB dominated scenario and---for low neutral fractions---the PL 3.0
scenario. A real-life measurement would give $\Ps$ as a function of
$z$, so if we assume that $\Ps(\xb)$ is known, we can use this to
reconstruct the reionization history, i.e.\ $\xb(z)$. Here, we attempt
to do this for our mock observations.

We begin by parametrizing $\xb(z)$ as follows:
\begin{align}
    \xb(z) = 
    \begin{cases} 0 & \text{if $z < z_r$} \\
		1 - \exp[-\beta(z-z_r)] &\text{if $z \geq z_r$}
\end{cases}
\label{eq:xh1_param}
\end{align}
The parameters $z_r$ and $\beta$ give the endpoint and extent of
reionization respectively (higher values of $\beta$ indicate a more
rapid reionization). Given values of $z_r$ and $\beta$ and a
measurement of $\Ps(z)$, we can use Equation \eqref{eq:xh1_param} to
construct $\Ps(\xb)$. Since we assume that this latter quantity is
known, we may tune $z_r$ and $\beta$ until the measured $\Ps(\xb)$
matches the true one.

\begin{figure*}
  \includegraphics[width=1.\textwidth]{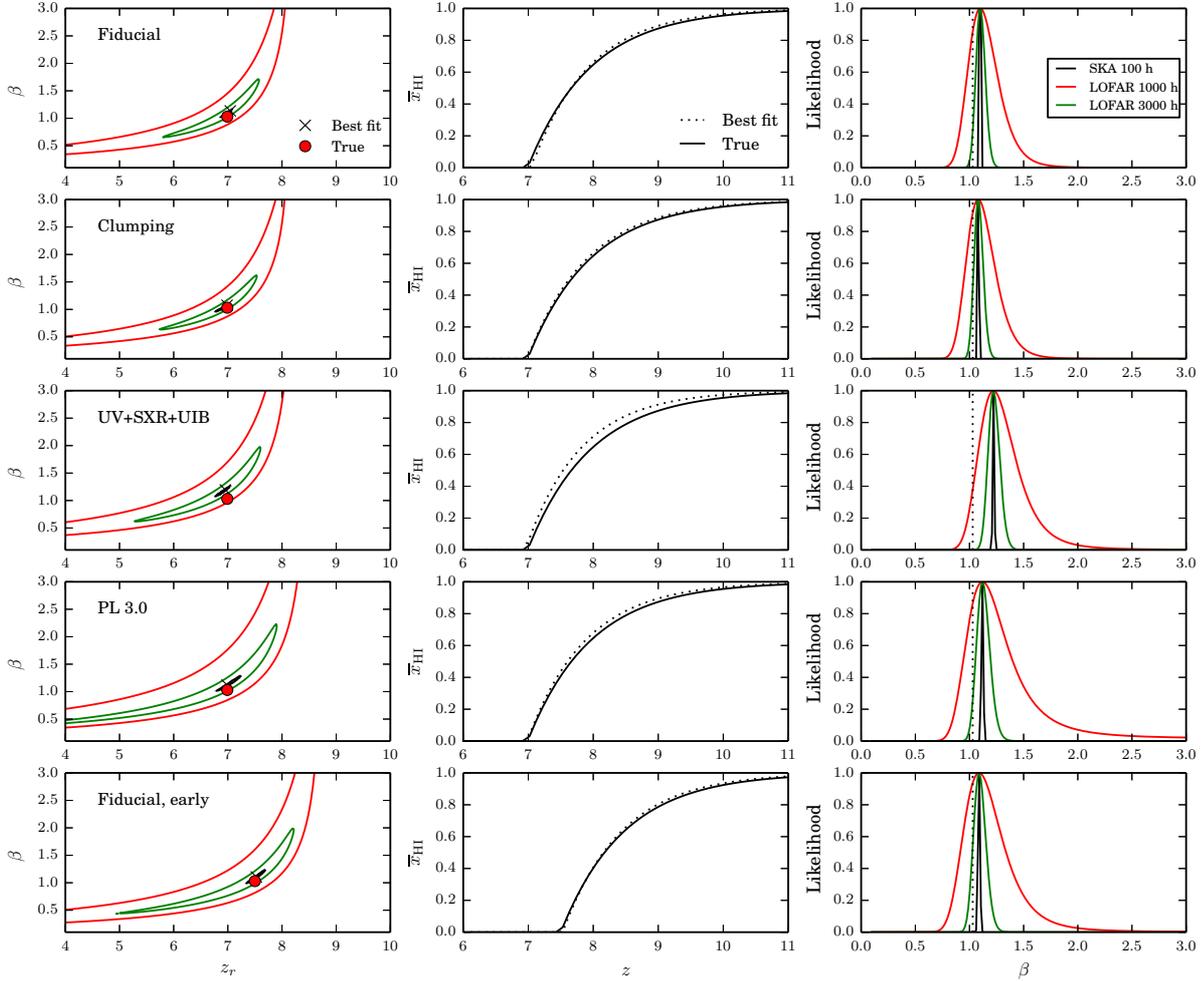}
  \caption{Fitting the reionization history to match the slope of
    $\Ps$ as a function of $\xb$. \emph{Left column}: Likelihood
    contours for 1000 and 3000 hours integration time with LOFAR and
    100 hours of integration time with SKA, for different reionization
    scenarios. The contours are located at the value where the
    likelihood is a factor $e^2$ lower than the maximum value,
    corresponding to roughly 90 per cent of the full distribution. The
    crosses show the best-fit values of the parameters $z_r$ and
    $\beta$ (in the absence of noise), and the red circles show the
    true values. \emph{Middle column}: The reionization histories
    corresponding to the best-fit values from the left column,
    compared to the true histories. \emph{Right column}: Same as the
    left column, but assuming that $z_r$ is known. The dotted vertical
    lines show the true values of $\beta$.}
  \label{fig:reconstructed}
\end{figure*}

If a measurement of $P_2^{\mathrm{s},\mathrm{meas}}(z)$ has Gaussian noise with
an amplitude of $\sigma$, the likelihood of measuring this value is:
\begin{equation}
  {\mathcal L} = \prod_i \frac{1}{\sqrt{2 \upi}\sigma}\exp \left[\frac{-\left( P_{2,i}^{s,\mathrm{meas}} - P_{2,i}^{s,\mathrm{true}}\right)^2}{2 \sigma^2}\right],
  \label{eq:likelihood}
\end{equation}
where the product is calculated over all redshifts where there are
measurements available. We are thus looking for the values of $z_r$
and $\beta$ that give rise to the $\xb$ history that maximizes the
likelihood.

Judging by Figure \ref{fig:p2}, $\Ps(\xb)$ can be assumed to be
reasonably well-known, especially in the early stages of reionization.
Here, we therefore only use the part of $\Ps$ where $\xb \ge 0.75$,
since the slope of $\Ps(\xb)$ is very consistent from scenario to
scenario. We take the values from the fiducial model lightcone to be
the ``true'' $\Ps(\xb)$. More specifically, we use a linear fit to the
values of $\Ps(\xb)$ for $\xb>0.75$ plotted in Figure \ref{fig:p2} to
be our $P_{2}^{s,\mathrm{true}}$.

Figure \ref{fig:reconstructed} shows the results of fitting the
reionization history. $P_2^{\mathrm{s,meas}}$ is here taken from our
simulated reionization scenarios (see Figure \ref{fig:p2}), and the
noise $\sigma$ is calculated as described in the previous section.  In
the left panel we show the likelihood contours for LOFAR noise for
1000 and 3000 hours of integration time and for SKA noise for 100
hours. The crosses show the parameter values that maximize the
likelihood, while the red circles show the true values, i.e.\ a fit of
Equation \eqref{eq:xh1_param} to the true reionization history (since
our simulations only go down to $\xb\sim0.2$, the value of $z_r$
represents an extrapolation; c.f.\ Figure \ref{fig:xh1mz}). The error
contours do not explicitly include the effects of sample
variance{\footnote{The inherent non-Gaussianity in the EoR 21-cm
    signal can make the cosmic variance at these length scales deviate
    from its generally assumed Gaussian behaviour
    \citep{mondal15,mondal15b}, increasing the uncertainties in the
    reconstruction of the EoR history.}}.  The middle panel of Figure
\ref{fig:reconstructed} shows the reconstructed reionization histories
corresponding to the best-fit parameters in the left panel.

The fact that the best-fit values are very close to the true values
for all reionization scenarios shows that the evolution of $\Ps$ can
indeed be used as a robust measurement of the reionization
history. For LOFAR, it is clear that noise is an important
obstacle. With 1000 hours of integration time, we can only put rather
loose constraints on the reionization history. After 3000 hours,
however, the constraints become much better.  For the SKA, the noise
is so low that even for 100 hours integration time, the reionization
history can be determined almost perfectly.

We also see that for our parametrization of the reionization history,
there is quite a bit of degeneracy between the two parameters, as seen
by the elongated likelihood contours. In a real-world scenario, it is
likely that there will be some information available on one of the
parameters. For example, $z_r$ may be estimated by looking for the
redshift when the spherically averaged power spectrum tends to
zero. In the right column of Figure \ref{fig:reconstructed}, we show
the results of fitting only $\beta$, assuming that $z_r$ is known. In
this case, LOFAR measurements may be able to say something about the
reionization history, or at least exclude meaningful regions of
parameter space.

While all of our reionization scenarios are constructed to have very
different topologies, they are also tuned to have the same
reionization history. A widely accepted view is that the 21-cm power
spectrum is much more sensitive to $\xb$ than $z$
(e.g.\ \citealt{iliev12}), and so we do not expect the reionization
history to have a large effect on our results. Nevertheless, to test
whether a different reionization history would change the shape of
$\Ps(\xb)$, we also constructed a scenario with the same source
properties as the fiducial scenario, but with the source efficiencies
increased, in order to produce an earlier reionization. The results
for fitting the reionization history for this scenario is shown in the
bottom row of Figure \ref{fig:reconstructed}, still using $\Ps(\xb)$
from the fiducial scenario as the true value. While a full
investigation of the effects of the exact reionization history is
beyond the scope of this paper, it is clear from this test that a
modest shift in the history has no significant effect on the results.

The analysis presented in this section is a proof-of-concept of
the possibility of extracting the reionization history from the
redshift-space anisotropy of the 21-cm signal. In actual
radio-interferometric observations, the presence of residuals from
foregrounds in the reduced 21-cm data can introduce further
uncertainties in these estimations, which we have not accounted for.

\section{Summary and conclusions}
\label{sec:summary}
The velocities of matter distort the 21-cm signal and cause the
power spectrum to become anisotropic. The anisotropy depends on the
cross-correlation between the density and the \HI fields and therefore
carries information both about the cosmology and about the properties
of the sources of reionization. In this paper we have explored a
number of different reionization scenarios with different types of
sources of ionizing photons in order to study the effect of the sources
of reionization on the power spectrum anisotropy.

We find that the power spectrum anisotropy, as measured by the
quadrupole moment of the power spectrum, $\Ps$, as a function of
global neutral fraction evolves in a very similar way for all
reionization scenarios in which the sources and the matter distribution
are strongly correlated. Only the most extreme scenarios, such as
reionization driven by a uniform ionizing background or by very
massive sources, show significant deviations from the others. Thus,
$\Ps$ is not a sensitive tool for separating reionization scenarios.
However, it can be used as a robust tracer of the
reionization history.

Interferometers such as LOFAR or the SKA can only measure the
fluctuations of the 21-cm signal, not the actual magnitude of the
signal. Therefore, extracting the reionization history is not trivial,
and can only be done by fitting a model to the observed data. Previous
studies have suggested that looking for the peak in the
spherically-averaged power spectrum (e.g.\
\citealt{mcquinn07,lidz08,barkana09,iliev12}) or the variance
\citep{patil14} will give the mid-point of reionization. However, as
seen in this paper and elsewhere (e.g.\
\citealt{mellema06,iliev06a,iliev07,watkinson15}), the
spherically-averaged power spectrum can in fact vary substantially
depending on the properties of the sources of reionization. The
quadrupole moment is much more model-independent, as long as the
reionization topology is inside-out. This is likely to be the case,
considering that the latest Planck results on the optical depth to
Thomson scattering combined with observations of high-redshift
galaxies strongly favour galaxies as the main driver of reionization
\citep{bouwens15}.

We have shown that measuring the reionization history using the
quadrupole moment works well for all reionization scenarios considered
in this paper, except for the most extreme ones. For first-generation
telescopes such as LOFAR, these types of measurements will be
challenging due to the uncertainties imposed by the noise and
foreground emission.  However, it should be possible to use the power
spectrum anisotropy to exclude certain reionization histories. In any
case the anisotropy can be used to verify that the observed signal is
indeed the 21-cm emission from the EoR.  For the SKA, neither the
noise uncertainty nor foregrounds will be an issue, and it will be
possible to accurately measure the reionization history even with a
relatively short integration time.

Of course, our treatment of observational effects here is simplified
and does not take into account complications from, for example,
the frequency-dependence of the PSF, ionospheric effects or calibration
errors. Nevertheless, it shows that the 21-cm power spectrum
anisotropy due to the redshift-space distortions has the potential to
be a powerful and nearly model-independent probe of the reionization
history.

\section*{Acknowledgements}
SM would like to thank Raghunath Ghara for useful discussions related
to the effect of spin temperature fluctuations on the EoR 21-cm
signal. The research described in this paper was supported by a grant
from the Lennart and Alva Dahlmark Fund. GM is supported by Swedish
Research Council project grant 2012-4144. The PRACE4LOFAR simulations
were performed on the Curie system at TGCC under PRACE projects
2012061089 and 2014102339. This work was supported by the Science and
Technology Facilities Council [grant number ST/L000652/1]. KKD would
like to thank DST for support through the project SR/FTP/PS-119/2012
and the University Grant Commission (UGC), India for support through
UGC-faculty recharge scheme (UGC-FRP) vide
ref. no. F.4-5(137-FRP)/2014(BSR). VJ would like to thank the
Netherlands Foundation for Scientific Research (NWO) for financial
support through VENI grant 639.041.336. LVEK acknowledges the
financial support from the European Research Council under
ERC-Starting Grant FIRSTLIGHT - 258942.

\bibliographystyle{mn2e} 
\bibliography{refs}

\begin{thebibliography}{86}
\expandafter\ifx\csname natexlab\endcsname\relax\def\natexlab#1{#1}\fi

\bibitem[{{Ali}, {Bharadwaj} \& {Chengalur}(2008){Ali}, {Bharadwaj}, \&
  {Chengalur}}]{ali08}
{Ali} S.~S., {Bharadwaj} S., {Chengalur} J.~N., 2008, \mnras, 385, 2166

\bibitem[{{Ali} {et~al}\mbox{.}(2015){Ali}, {Parsons}, {Zheng}, {Pober}, {Liu},
  {Aguirre}, {Bradley}, {Bernardi}, {Carilli}, {Cheng}, {DeBoer}, {Dexter},
  {Grobbelaar}, {Horrell}, {Jacobs}, {Klima}, {MacMahon}, {Maree}, {Moore},
  {Razavi}, {Stefan}, {Walbrugh}, \& {Walker}}]{ali15}
{Ali} Z.~S. {et~al.}, 2015, \apj, 809, 61

\bibitem[{{Alvarez} {et~al}\mbox{.}(2006){Alvarez}, {Shapiro}, {Ahn}, \&
  {Iliev}}]{alvarez06}
{Alvarez} M.~A., {Shapiro} P.~R., {Ahn} K., {Iliev} I.~T., 2006, \apjl, 644,
  L101

\bibitem[{{Barkana}(2009)}]{barkana09}
{Barkana} R., 2009, \mnras, 397, 1454

\bibitem[{{Barkana} \& {Loeb}(2005)}]{barkana05}
{Barkana} R., {Loeb} A., 2005, \apjl, 624, L65

\bibitem[{{Becker} {et~al}\mbox{.}(2015){Becker}, {Bolton}, {Madau}, {Pettini},
  {Ryan-Weber}, \& {Venemans}}]{becker15}
{Becker} G.~D., {Bolton} J.~S., {Madau} P., {Pettini} M., {Ryan-Weber} E.~V.,
  {Venemans} B.~P., 2015, \mnras, 447, 3402

\bibitem[{{Becker} {et~al}\mbox{.}(2001){Becker}, {Fan}, {White}, {Strauss},
  {Narayanan}, {Lupton}, {Gunn}, {Annis}, {Bahcall}, {Brinkmann}, {Connolly},
  {Csabai}, {Czarapata}, {Doi}, {Heckman}, {Hennessy}, {Ivezi{\'c}}, {Knapp},
  {Lamb}, {McKay}, {Munn}, {Nash}, {Nichol}, {Pier}, {Richards}, {Schneider},
  {Stoughton}, {Szalay}, {Thakar}, \& {York}}]{becker01}
{Becker} R.~H. {et~al.}, 2001, \aj, 122, 2850

\bibitem[{{Bharadwaj} \& {Ali}(2004)}]{bharadwaj04}
{Bharadwaj} S., {Ali} S.~S., 2004, \mnras, 352, 142

\bibitem[{{Bharadwaj} \& {Ali}(2005)}]{bharadwaj05}
{Bharadwaj} S., {Ali} S.~S., 2005, \mnras, 356, 1519

\bibitem[{{Bharadwaj} \& {Pandey}(2005)}]{bharadwaj05a}
{Bharadwaj} S., {Pandey} S.~K., 2005, \mnras, 358, 968

\bibitem[{{Bouwens} {et~al}\mbox{.}(2015){Bouwens}, {Illingworth}, {Oesch},
  {Caruana}, {Holwerda}, {Smit}, \& {Wilkins}}]{bouwens15}
{Bouwens} R.~J., {Illingworth} G.~D., {Oesch} P.~A., {Caruana} J., {Holwerda}
  B., {Smit} R., {Wilkins} S., 2015, ArXiv e-prints; arXiv:1503.08228

\bibitem[{{Bowman} {et~al}\mbox{.}(2013){Bowman}, {Cairns}, {Kaplan}, {Murphy},
  {Oberoi}, {Staveley-Smith}, {Arcus}, {Barnes}, {Bernardi}, {Briggs}, {Brown},
  {Bunton}, {Burgasser}, {Cappallo}, {Chatterjee}, {Corey}, {Coster},
  {Deshpande}, {deSouza}, {Emrich}, {Erickson}, {Goeke}, {Gaensler},
  {Greenhill}, {Harvey-Smith}, {Hazelton}, {Herne}, {Hewitt},
  {Johnston-Hollitt}, {Kasper}, {Kincaid}, {Koenig}, {Kratzenberg}, {Lonsdale},
  {Lynch}, {Matthews}, {McWhirter}, {Mitchell}, {Morales}, {Morgan}, {Ord},
  {Pathikulangara}, {Prabu}, {Remillard}, {Robishaw}, {Rogers}, {Roshi},
  {Salah}, {Sault}, {Shankar}, {Srivani}, {Stevens}, {Subrahmanyan}, {Tingay},
  {Wayth}, {Waterson}, {Webster}, {Whitney}, {Williams}, {Williams}, \&
  {Wyithe}}]{bowman13}
{Bowman} J.~D. {et~al.}, 2013, \pasa, 30, 31

\bibitem[{{Chapman} {et~al}\mbox{.}(2013){Chapman}, {Abdalla}, {Bobin},
  {Starck}, {Harker}, {Jeli{\'c}}, {Labropoulos}, {Zaroubi}, {Brentjens}, {de
  Bruyn}, \& {Koopmans}}]{chapman2013}
{Chapman} E. {et~al.}, 2013, \mnras, 429, 165

\bibitem[{{Choudhury}, {Haehnelt} \& {Regan}(2009){Choudhury}, {Haehnelt}, \&
  {Regan}}]{choudhury09b}
{Choudhury} T.~R., {Haehnelt} M.~G., {Regan} J., 2009, \mnras, 394, 960

\bibitem[{{Choudhury} {et~al}\mbox{.}(2015){Choudhury}, {Puchwein}, {Haehnelt},
  \& {Bolton}}]{choudhury14}
{Choudhury} T.~R., {Puchwein} E., {Haehnelt} M.~G., {Bolton} J.~S., 2015,
  \mnras, 452, 261

\bibitem[{{Cole}, {Fisher} \& {Weinberg}(1995){Cole}, {Fisher}, \&
  {Weinberg}}]{cole95}
{Cole} S., {Fisher} K.~B., {Weinberg} D.~H., 1995, \mnras, 275, 515

\bibitem[{{Couchman} \& {Rees}(1986)}]{couchman86}
{Couchman} H.~M.~P., {Rees} M.~J., 1986, \mnras, 221, 53

\bibitem[{{Datta} {et~al}\mbox{.}(2014){Datta}, {Jensen}, {Majumdar},
  {Mellema}, {Iliev}, {Mao}, {Shapiro}, \& {Ahn}}]{datta14}
{Datta} K.~K., {Jensen} H., {Majumdar} S., {Mellema} G., {Iliev} I.~T., {Mao}
  Y., {Shapiro} P.~R., {Ahn} K., 2014, \mnras, 442, 1491

\bibitem[{{Datta} {et~al}\mbox{.}(2012){Datta}, {Mellema}, {Mao}, {Iliev},
  {Shapiro}, \& {Ahn}}]{datta12}
{Datta} K.~K., {Mellema} G., {Mao} Y., {Iliev} I.~T., {Shapiro} P.~R., {Ahn}
  K., 2012, \mnras, 424, 1877

\bibitem[{{Dewdney} {et~al}\mbox{.}(2013){Dewdney}, {Turner}, {Millenaar},
  {McCool}, {Lazio}, \& {Cornwell}}]{dewdney13}
{Dewdney} P., {Turner} W., {Millenaar} R., {McCool} R., {Lazio} J., {Cornwell}
  T., 2013, SKA1 System Baseline Design, Document number SKA-TEL-SKO-DD-001
  Revision 1, 1

\bibitem[{{Di Matteo} {et~al}\mbox{.}(2002){Di Matteo}, {Perna}, {Abel}, \&
  {Rees}}]{dimatteo02}
{Di Matteo} T., {Perna} R., {Abel} T., {Rees} M.~J., 2002, \apj, 564, 576

\bibitem[{{Dijkstra} {et~al}\mbox{.}(2004){Dijkstra}, {Haiman}, {Rees}, \&
  {Weinberg}}]{dijkstra04}
{Dijkstra} M., {Haiman} Z., {Rees} M.~J., {Weinberg} D.~H., 2004, \apj, 601,
  666

\bibitem[{{Dillon} {et~al}\mbox{.}(2014){Dillon}, {Liu}, {Williams}, {Hewitt},
  {Tegmark}, {Morgan}, {Levine}, {Morales}, {Tingay}, {Bernardi}, {Bowman},
  {Briggs}, {Cappallo}, {Emrich}, {Mitchell}, {Oberoi}, {Prabu}, {Wayth}, \&
  {Webster}}]{dillon14}
{Dillon} J.~S. {et~al.}, 2014, \prd, 89, 023002

\bibitem[{{Fan} {et~al}\mbox{.}(2003){Fan}, {Strauss}, {Schneider}, {Becker},
  {White}, {Haiman}, {Gregg}, {Pentericci}, {Grebel}, {Narayanan}, {Loh},
  {Richards}, {Gunn}, {Lupton}, {Knapp}, {Ivezi{\'c}}, {Brandt}, {Collinge},
  {Hao}, {Harbeck}, {Prada}, {Schaye}, {Strateva}, {Zakamska}, {Anderson},
  {Brinkmann}, {Bahcall}, {Lamb}, {Okamura}, {Szalay}, \& {York}}]{fan03}
{Fan} X. {et~al.}, 2003, \aj, 125, 1649

\bibitem[{{Fialkov}, {Barkana} \& {Cohen}(2015){Fialkov}, {Barkana}, \&
  {Cohen}}]{fialkov15}
{Fialkov} A., {Barkana} R., {Cohen} A., 2015, Physical Review Letters, 114,
  101303

\bibitem[{{Fialkov}, {Barkana} \& {Visbal}(2014){Fialkov}, {Barkana}, \&
  {Visbal}}]{fialkov14a}
{Fialkov} A., {Barkana} R., {Visbal} E., 2014, \nat, 506, 197

\bibitem[{{Fragos} {et~al}\mbox{.}(2013){Fragos}, {Lehmer}, {Tremmel},
  {Tzanavaris}, {Basu-Zych}, {Belczynski}, {Hornschemeier}, {Jenkins},
  {Kalogera}, {Ptak}, \& {Zezas}}]{fragos13}
{Fragos} T. {et~al.}, 2013, \apj, 764, 41

\bibitem[{{Furlanetto}, {Zaldarriaga} \& {Hernquist}(2004){Furlanetto},
  {Zaldarriaga}, \& {Hernquist}}]{furlanetto04b}
{Furlanetto} S.~R., {Zaldarriaga} M., {Hernquist} L., 2004, \apj, 613, 1

\bibitem[{{Ghara}, {Choudhury} \& {Datta}(2015){Ghara}, {Choudhury}, \&
  {Datta}}]{ghara14}
{Ghara} R., {Choudhury} T.~R., {Datta} K.~K., 2015, \mnras, 447, 1806

\bibitem[{{Ghara}, {Datta} \& {Choudhury}(2015){Ghara}, {Datta}, \&
  {Choudhury}}]{ghara15}
{Ghara} R., {Datta} K.~K., {Choudhury} T.~R., 2015, \mnras, 453, 3143

\bibitem[{{Gnedin}(2000)}]{gnedin00b}
{Gnedin} N.~Y., 2000, \apj, 542, 535

\bibitem[{{Goto} {et~al}\mbox{.}(2011){Goto}, {Utsumi}, {Hattori}, {Miyazaki},
  \& {Yamauchi}}]{goto11}
{Goto} T., {Utsumi} Y., {Hattori} T., {Miyazaki} S., {Yamauchi} C., 2011,
  \mnras, 415, L1

\bibitem[{{Hamilton}(1992)}]{hamilton92}
{Hamilton} A.~J.~S., 1992, \apjl, 385, L5

\bibitem[{{Hamilton}(1998)}]{hamilton98}
{Hamilton} A.~J.~S., 1998, in Astrophysics and Space Science Library, Vol. 231,
  The Evolving Universe, {Hamilton} D., ed., p. 185

\bibitem[{{Harnois-D{\'e}raps} {et~al}\mbox{.}(2013){Harnois-D{\'e}raps},
  {Pen}, {Iliev}, {Merz}, {Emberson}, \& {Desjacques}}]{harnoisderaps13}
{Harnois-D{\'e}raps} J., {Pen} U.-L., {Iliev} I.~T., {Merz} H., {Emberson}
  J.~D., {Desjacques} V., 2013, \mnras, 436, 540

\bibitem[{{Iliev} {et~al}\mbox{.}(2006){Iliev}, {Mellema}, {Pen}, {Merz},
  {Shapiro}, \& {Alvarez}}]{iliev06a}
{Iliev} I.~T., {Mellema} G., {Pen} U.-L., {Merz} H., {Shapiro} P.~R., {Alvarez}
  M.~A., 2006, \mnras, 369, 1625

\bibitem[{{Iliev} {et~al}\mbox{.}(2007){Iliev}, {Mellema}, {Shapiro}, \&
  {Pen}}]{iliev07}
{Iliev} I.~T., {Mellema} G., {Shapiro} P.~R., {Pen} U.-L., 2007, \mnras, 376,
  534

\bibitem[{{Iliev} {et~al}\mbox{.}(2012){Iliev}, {Mellema}, {Shapiro}, {Pen},
  {Mao}, {Koda}, \& {Ahn}}]{iliev12}
{Iliev} I.~T., {Mellema} G., {Shapiro} P.~R., {Pen} U.-L., {Mao} Y., {Koda} J.,
  {Ahn} K., 2012, \mnras, 423, 2222

\bibitem[{{Jeli{\'c}} {et~al}\mbox{.}(2014){Jeli{\'c}}, {de Bruyn}, {Mevius},
  {Abdalla}, {Asad}, {Bernardi}, {Brentjens}, {Bus}, {Chapman}, {Ciardi},
  {Daiboo}, {Fernandez}, {Ghosh}, {Harker}, {Jensen}, {Kazemi}, {Koopmans},
  {Labropoulos}, {Martinez-Rubi}, {Mellema}, {Offringa}, {Pandey}, {Patil},
  {Thomas}, {Vedantham}, {Veligatla}, {Yatawatta}, {Zaroubi}, {Alexov},
  {Anderson}, {Avruch}, {Beck}, {Bell}, {Bentum}, {Best}, {Bonafede},
  {Bregman}, {Breitling}, {Broderick}, {Brouw}, {Br{\"u}ggen}, {Butcher},
  {Conway}, {de Gasperin}, {de Geus}, {Deller}, {Dettmar}, {Duscha},
  {Eisl{\"o}ffel}, {Engels}, {Falcke}, {Fallows}, {Fender}, {Ferrari},
  {Frieswijk}, {Garrett}, {Grie{\ss}meier}, {Gunst}, {Hamaker}, {Hassall},
  {Haverkorn}, {Heald}, {Hessels}, {Hoeft}, {H{\"o}randel}, {Horneffer}, {van
  der Horst}, {Iacobelli}, {Juette}, {Karastergiou}, {Kondratiev}, {Kramer},
  {Kuniyoshi}, {Kuper}, {van Leeuwen}, {Maat}, {Mann}, {McKay-Bukowski},
  {McKean}, {Munk}, {Nelles}, {Norden}, {Paas}, {Pandey-Pommier}, {Pietka},
  {Pizzo}, {Polatidis}, {Reich}, {R{\"o}ttgering}, {Rowlinson}, {Scaife},
  {Schwarz}, {Serylak}, {Smirnov}, {Steinmetz}, {Stewart}, {Tagger}, {Tang},
  {Tasse}, {ter Veen}, {Thoudam}, {Toribio}, {Vermeulen}, {Vocks}, {van
  Weeren}, {Wijers}, {Wijnholds}, {Wucknitz}, \& {Zarka}}]{jelic14}
{Jeli{\'c}} V. {et~al.}, 2014, \aap, 568, A101

\bibitem[{{Jeli{\'c}} {et~al}\mbox{.}(2010){Jeli{\'c}}, {Zaroubi},
  {Labropoulos}, {Bernardi}, {de Bruyn}, \& {Koopmans}}]{jelic10}
{Jeli{\'c}} V., {Zaroubi} S., {Labropoulos} P., {Bernardi} G., {de Bruyn}
  A.~G., {Koopmans} L.~V.~E., 2010, \mnras, 409, 1647

\bibitem[{{Jeli{\'c}} {et~al}\mbox{.}(2008){Jeli{\'c}}, {Zaroubi},
  {Labropoulos}, {Thomas}, {Bernardi}, {Brentjens}, {de Bruyn}, {Ciardi},
  {Harker}, {Koopmans}, {Pandey}, {Schaye}, \& {Yatawatta}}]{jelic08}
{Jeli{\'c}} V. {et~al.}, 2008, \mnras, 389, 1319

\bibitem[{{Jensen} {et~al}\mbox{.}(2013){Jensen}, {Datta}, {Mellema},
  {Chapman}, {Abdalla}, {Iliev}, {Mao}, {Santos}, {Shapiro}, {Zaroubi},
  {Bernardi}, {Brentjens}, {de Bruyn}, {Ciardi}, {Harker}, {Jeli{\'c}},
  {Kazemi}, {Koopmans}, {Labropoulos}, {Martinez}, {Offringa}, {Pandey},
  {Schaye}, {Thomas}, {Veligatla}, {Vedantham}, \& {Yatawatta}}]{jensen13}
{Jensen} H. {et~al.}, 2013, \mnras, 435, 460

\bibitem[{{Komatsu} {et~al}\mbox{.}(2009){Komatsu}, {Dunkley}, {Nolta},
  {Bennett}, {Gold}, {Hinshaw}, {Jarosik}, {Larson}, {Limon}, {Page},
  {Spergel}, {Halpern}, {Hill}, {Kogut}, {Meyer}, {Tucker}, {Weiland},
  {Wollack}, \& {Wright}}]{komatsu09}
{Komatsu} E. {et~al.}, 2009, \apjs, 180, 330

\bibitem[{{Komatsu} {et~al}\mbox{.}(2011){Komatsu}, {Smith}, {Dunkley},
  {Bennett}, {Gold}, {Hinshaw}, {Jarosik}, {Larson}, {Nolta}, {Page},
  {Spergel}, {Halpern}, {Hill}, {Kogut}, {Limon}, {Meyer}, {Odegard}, {Tucker},
  {Weiland}, {Wollack}, \& {Wright}}]{komatsu11}
{Komatsu} E. {et~al.}, 2011, \apjs, 192, 18

\bibitem[{{Koopmans} {et~al}\mbox{.}(2015){Koopmans}, {Pritchard}, {Mellema},
  {Aguirre}, {Ahn}, {Barkana}, {van Bemmel}, {Bernardi}, {Bonaldi}, {Briggs},
  {de Bruyn}, {Chang}, {Chapman}, {Chen}, {Ciardi}, {Dayal}, {Ferrara},
  {Fialkov}, {Fiore}, {Ichiki}, {Illiev}, {Inoue}, {Jelic}, {Jones}, {Lazio},
  {Maio}, {Majumdar}, {Mack}, {Mesinger}, {Morales}, {Parsons}, {Pen},
  {Santos}, {Schneider}, {Semelin}, {de Souza}, {Subrahmanyan}, {Takeuchi},
  {Vedantham}, {Wagg}, {Webster}, {Wyithe}, {Datta}, \& {Trott}}]{koopmans15}
{Koopmans} L. {et~al.}, 2015, Advancing Astrophysics with the Square Kilometre
  Array (AASKA14), 1

\bibitem[{{Kuhlen} \& {Faucher-Gigu{\`e}re}(2012)}]{kuhlen12}
{Kuhlen} M., {Faucher-Gigu{\`e}re} C.-A., 2012, \mnras, 423, 862

\bibitem[{{Lidz} {et~al}\mbox{.}(2008){Lidz}, {Zahn}, {McQuinn}, {Zaldarriaga},
  \& {Hernquist}}]{lidz08}
{Lidz} A., {Zahn} O., {McQuinn} M., {Zaldarriaga} M., {Hernquist} L., 2008,
  \apj, 680, 962

\bibitem[{{Madau}, {Haardt} \& {Rees}(1999){Madau}, {Haardt}, \&
  {Rees}}]{madau98}
{Madau} P., {Haardt} F., {Rees} M.~J., 1999, \apj, 514, 648

\bibitem[{{Majumdar}, {Bharadwaj} \& {Choudhury}(2013){Majumdar}, {Bharadwaj},
  \& {Choudhury}}]{majumdar13}
{Majumdar} S., {Bharadwaj} S., {Choudhury} T.~R., 2013, \mnras, 434, 1978

\bibitem[{{Majumdar} {et~al}\mbox{.}(2014){Majumdar}, {Mellema}, {Datta},
  {Jensen}, {Choudhury}, {Bharadwaj}, \& {Friedrich}}]{majumdar14}
{Majumdar} S., {Mellema} G., {Datta} K.~K., {Jensen} H., {Choudhury} T.~R.,
  {Bharadwaj} S., {Friedrich} M.~M., 2014, \mnras, 443, 2843

\bibitem[{{Mao} {et~al}\mbox{.}(2012){Mao}, {Shapiro}, {Mellema}, {Iliev},
  {Koda}, \& {Ahn}}]{mao12}
{Mao} Y., {Shapiro} P.~R., {Mellema} G., {Iliev} I.~T., {Koda} J., {Ahn} K.,
  2012, \mnras, 422, 926

\bibitem[{{McQuinn}(2012)}]{mcquinn12}
{McQuinn} M., 2012, \mnras, 426, 1349

\bibitem[{{McQuinn} {et~al}\mbox{.}(2007){McQuinn}, {Lidz}, {Zahn}, {Dutta},
  {Hernquist}, \& {Zaldarriaga}}]{mcquinn07}
{McQuinn} M., {Lidz} A., {Zahn} O., {Dutta} S., {Hernquist} L., {Zaldarriaga}
  M., 2007, \mnras, 377, 1043

\bibitem[{{McQuinn} {et~al}\mbox{.}(2006){McQuinn}, {Zahn}, {Zaldarriaga},
  {Hernquist}, \& {Furlanetto}}]{mcquinn06}
{McQuinn} M., {Zahn} O., {Zaldarriaga} M., {Hernquist} L., {Furlanetto} S.~R.,
  2006, \apj, 653, 815

\bibitem[{{Mellema} {et~al}\mbox{.}(2006){Mellema}, {Iliev}, {Pen}, \&
  {Shapiro}}]{mellema06}
{Mellema} G., {Iliev} I.~T., {Pen} U.-L., {Shapiro} P.~R., 2006, \mnras, 372,
  679

\bibitem[{{Mellema} {et~al}\mbox{.}(2015){Mellema}, {Koopmans}, {Shukla},
  {Datta}, {Mesinger}, \& {Majumdar}}]{mellema15}
{Mellema} G., {Koopmans} L., {Shukla} H., {Datta} K.~K., {Mesinger} A.,
  {Majumdar} S., 2015, Advancing Astrophysics with the Square Kilometre Array
  (AASKA14), 10

\bibitem[{{Mellema} {et~al}\mbox{.}(2013){Mellema}, {Koopmans}, {Abdalla},
  {Bernardi}, {Ciardi}, {Daiboo}, {de Bruyn}, {Datta}, {Falcke}, {Ferrara},
  {Iliev}, {Iocco}, {Jeli{\'c}}, {Jensen}, {Joseph}, {Labroupoulos}, {Meiksin},
  {Mesinger}, {Offringa}, {Pandey}, {Pritchard}, {Santos}, {Schwarz},
  {Semelin}, {Vedantham}, {Yatawatta}, \& {Zaroubi}}]{mellema13}
{Mellema} G. {et~al.}, 2013, Experimental Astronomy, 36, 235

\bibitem[{{Merz}, {Pen} \& {Trac}(2005){Merz}, {Pen}, \& {Trac}}]{merz05}
{Merz} H., {Pen} U.-L., {Trac} H., 2005, \na, 10, 393

\bibitem[{{Mesinger}, {Ferrara} \& {Spiegel}(2013){Mesinger}, {Ferrara}, \&
  {Spiegel}}]{mesinger13}
{Mesinger} A., {Ferrara} A., {Spiegel} D.~S., 2013, \mnras, 431, 621

\bibitem[{{Mesinger} \& {Furlanetto}(2007)}]{mesinger07}
{Mesinger} A., {Furlanetto} S., 2007, \apj, 669, 663

\bibitem[{{Mesinger}, {Furlanetto} \& {Cen}(2011){Mesinger}, {Furlanetto}, \&
  {Cen}}]{mesinger11}
{Mesinger} A., {Furlanetto} S., {Cen} R., 2011, \mnras, 411, 955

\bibitem[{{Mirabel} {et~al}\mbox{.}(2011){Mirabel}, {Dijkstra}, {Laurent},
  {Loeb}, \& {Pritchard}}]{mirabel11}
{Mirabel} I.~F., {Dijkstra} M., {Laurent} P., {Loeb} A., {Pritchard} J.~R.,
  2011, \aap, 528, A149

\bibitem[{{Mitra}, {Choudhury} \& {Ferrara}(2015){Mitra}, {Choudhury}, \&
  {Ferrara}}]{mitra15}
{Mitra} S., {Choudhury} T.~R., {Ferrara} A., 2015, ArXiv e-prints,
  arXiv:1505.05507

\bibitem[{{Mitra}, {Ferrara} \& {Choudhury}(2013){Mitra}, {Ferrara}, \&
  {Choudhury}}]{mitra13}
{Mitra} S., {Ferrara} A., {Choudhury} T.~R., 2013, \mnras, 428, L1

\bibitem[{{Mondal}, {Bharadwaj} \& {Majumdar}(2015){Mondal}, {Bharadwaj}, \&
  {Majumdar}}]{mondal15b}
{Mondal} R., {Bharadwaj} S., {Majumdar} S., 2015, ArXiv e-prints,
  arXiv:1508.00896

\bibitem[{{Mondal} {et~al}\mbox{.}(2015){Mondal}, {Bharadwaj}, {Majumdar},
  {Bera}, \& {Acharyya}}]{mondal15}
{Mondal} R., {Bharadwaj} S., {Majumdar} S., {Bera} A., {Acharyya} A., 2015,
  \mnras, 449, L41

\bibitem[{{Morales}(2005)}]{morales05}
{Morales} M.~F., 2005, \apj, 619, 678

\bibitem[{{Okamoto}, {Gao} \& {Theuns}(2008){Okamoto}, {Gao}, \&
  {Theuns}}]{okamoto08}
{Okamoto} T., {Gao} L., {Theuns} T., 2008, \mnras, 390, 920

\bibitem[{{Paciga} {et~al}\mbox{.}(2013){Paciga}, {Albert}, {Bandura}, {Chang},
  {Gupta}, {Hirata}, {Odegova}, {Pen}, {Peterson}, {Roy}, {Shaw}, {Sigurdson},
  \& {Voytek}}]{paciga13}
{Paciga} G. {et~al.}, 2013, \mnras, 433, 639

\bibitem[{{Parsons} {et~al}\mbox{.}(2014){Parsons}, {Liu}, {Aguirre}, {Ali},
  {Bradley}, {Carilli}, {DeBoer}, {Dexter}, {Gugliucci}, {Jacobs}, {Klima},
  {MacMahon}, {Manley}, {Moore}, {Pober}, {Stefan}, \& {Walbrugh}}]{parsons14}
{Parsons} A.~R. {et~al.}, 2014, \apj, 788, 106

\bibitem[{{Patil} {et~al}\mbox{.}(2014){Patil}, {Zaroubi}, {Chapman},
  {Jeli{\'c}}, {Harker}, {Abdalla}, {Asad}, {Bernardi}, {Brentjens}, {de
  Bruyn}, {Bus}, {Ciardi}, {Daiboo}, {Fernandez}, {Ghosh}, {Jensen}, {Kazemi},
  {Koopmans}, {Labropoulos}, {Mevius}, {Martinez}, {Mellema}, {Offringa},
  {Pandey}, {Schaye}, {Thomas}, {Vedantham}, {Veligatla}, {Wijnholds}, \&
  {Yatawatta}}]{patil14}
{Patil} A.~H. {et~al.}, 2014, \mnras, 443, 1113

\bibitem[{{Planck Collaboration}(2015)}]{planck15}
{Planck Collaboration}, 2015, ArXiv e-prints, arXiv:1502.01589

\bibitem[{{Pober} {et~al}\mbox{.}(2014){Pober}, {Liu}, {Dillon}, {Aguirre},
  {Bowman}, {Bradley}, {Carilli}, {DeBoer}, {Hewitt}, {Jacobs}, {McQuinn},
  {Morales}, {Parsons}, {Tegmark}, \& {Werthimer}}]{pober14}
{Pober} J.~C. {et~al.}, 2014, \apj, 782, 66

\bibitem[{{Robertson} {et~al}\mbox{.}(2015){Robertson}, {Ellis}, {Furlanetto},
  \& {Dunlop}}]{robertson15}
{Robertson} B.~E., {Ellis} R.~S., {Furlanetto} S.~R., {Dunlop} J.~S., 2015,
  \apjl, 802, L19

\bibitem[{{Santos} {et~al}\mbox{.}(2010){Santos}, {Ferramacho}, {Silva},
  {Amblard}, \& {Cooray}}]{santos10}
{Santos} M.~G., {Ferramacho} L., {Silva} M.~B., {Amblard} A., {Cooray} A.,
  2010, \mnras, 406, 2421

\bibitem[{{Shapiro} {et~al}\mbox{.}(2013){Shapiro}, {Mao}, {Iliev}, {Mellema},
  {Datta}, {Ahn}, \& {Koda}}]{shapiro13}
{Shapiro} P.~R., {Mao} Y., {Iliev} I.~T., {Mellema} G., {Datta} K.~K., {Ahn}
  K., {Koda} J., 2013, Physical Review Letters, 110, 151301

\bibitem[{{Sobacchi} \& {Mesinger}(2014)}]{sobacchi14}
{Sobacchi} E., {Mesinger} A., 2014, \mnras, 440, 1662

\bibitem[{{Songaila} \& {Cowie}(2010)}]{songaila10}
{Songaila} A., {Cowie} L.~L., 2010, \apj, 721, 1448

\bibitem[{{Tingay} {et~al}\mbox{.}(2013){Tingay}, {Goeke}, {Bowman}, {Emrich},
  {Ord}, {Mitchell}, {Morales}, {Booler}, {Crosse}, {Wayth}, {Lonsdale},
  {Tremblay}, {Pallot}, {Colegate}, {Wicenec}, {Kudryavtseva}, {Arcus},
  {Barnes}, {Bernardi}, {Briggs}, {Burns}, {Bunton}, {Cappallo}, {Corey},
  {Deshpande}, {Desouza}, {Gaensler}, {Greenhill}, {Hall}, {Hazelton}, {Herne},
  {Hewitt}, {Johnston-Hollitt}, {Kaplan}, {Kasper}, {Kincaid}, {Koenig},
  {Kratzenberg}, {Lynch}, {Mckinley}, {Mcwhirter}, {Morgan}, {Oberoi},
  {Pathikulangara}, {Prabu}, {Remillard}, {Rogers}, {Roshi}, {Salah}, {Sault},
  {Udaya-Shankar}, {Schlagenhaufer}, {Srivani}, {Stevens}, {Subrahmanyan},
  {Waterson}, {Webster}, {Whitney}, {Williams}, {Williams}, \&
  {Wyithe}}]{tingay13}
{Tingay} S.~J. {et~al.}, 2013, \pasa, 30, 7

\bibitem[{{van Haarlem} {et~al}\mbox{.}(2013){van Haarlem}, {Wise}, {Gunst},
  {Heald}, {McKean}, {Hessels}, {de Bruyn}, {Nijboer}, {Swinbank}, {Fallows},
  {Brentjens}, {Nelles}, {Beck}, {Falcke}, {Fender}, {H{\"o}randel},
  {Koopmans}, {Mann}, {Miley}, {R{\"o}ttgering}, {Stappers}, {Wijers},
  {Zaroubi}, {van den Akker}, {Alexov}, {Anderson}, {Anderson}, {van Ardenne},
  {Arts}, {Asgekar}, {Avruch}, {Batejat}, {B{\"a}hren}, {Bell}, {Bell}, {van
  Bemmel}, {Bennema}, {Bentum}, {Bernardi}, {Best}, {B{\^i}rzan}, {Bonafede},
  {Boonstra}, {Braun}, {Bregman}, {Breitling}, {van de Brink}, {Broderick},
  {Broekema}, {Brouw}, {Br{\"u}ggen}, {Butcher}, {van Cappellen}, {Ciardi},
  {Coenen}, {Conway}, {Coolen}, {Corstanje}, {Damstra}, {Davies}, {Deller},
  {Dettmar}, {van Diepen}, {Dijkstra}, {Donker}, {Doorduin}, {Dromer}, {Drost},
  {van Duin}, {Eisl{\"o}ffel}, {van Enst}, {Ferrari}, {Frieswijk}, {Gankema},
  {Garrett}, {de Gasperin}, {Gerbers}, {de Geus}, {Grie{\ss}meier}, {Grit},
  {Gruppen}, {Hamaker}, {Hassall}, {Hoeft}, {Holties}, {Horneffer}, {van der
  Horst}, {van Houwelingen}, {Huijgen}, {Iacobelli}, {Intema}, {Jackson},
  {Jelic}, {de Jong}, {Juette}, {Kant}, {Karastergiou}, {Koers}, {Kollen},
  {Kondratiev}, {Kooistra}, {Koopman}, {Koster}, {Kuniyoshi}, {Kramer},
  {Kuper}, {Lambropoulos}, {Law}, {van Leeuwen}, {Lemaitre}, {Loose}, {Maat},
  {Macario}, {Markoff}, {Masters}, {McFadden}, {McKay-Bukowski}, {Meijering},
  {Meulman}, {Mevius}, {Middelberg}, {Millenaar}, {Miller-Jones}, {Mohan},
  {Mol}, {Morawietz}, {Morganti}, {Mulcahy}, {Mulder}, {Munk}, {Nieuwenhuis},
  {van Nieuwpoort}, {Noordam}, {Norden}, {Noutsos}, {Offringa}, {Olofsson},
  {Omar}, {Orr{\'u}}, {Overeem}, {Paas}, {Pandey-Pommier}, {Pandey}, {Pizzo},
  {Polatidis}, {Rafferty}, {Rawlings}, {Reich}, {de Reijer}, {Reitsma},
  {Renting}, {Riemers}, {Rol}, {Romein}, {Roosjen}, {Ruiter}, {Scaife}, {van
  der Schaaf}, {Scheers}, {Schellart}, {Schoenmakers}, {Schoonderbeek},
  {Serylak}, {Shulevski}, {Sluman}, {Smirnov}, {Sobey}, {Spreeuw}, {Steinmetz},
  {Sterks}, {Stiepel}, {Stuurwold}, {Tagger}, {Tang}, {Tasse}, {Thomas},
  {Thoudam}, {Toribio}, {van der Tol}, {Usov}, {van Veelen}, {van der Veen},
  {ter Veen}, {Verbiest}, {Vermeulen}, {Vermaas}, {Vocks}, {Vogt}, {de Vos},
  {van der Wal}, {van Weeren}, {Weggemans}, {Weltevrede}, {White}, {Wijnholds},
  {Wilhelmsson}, {Wucknitz}, {Yatawatta}, {Zarka}, {Zensus}, \& {van
  Zwieten}}]{haarlem13}
{van Haarlem} M.~P. {et~al.}, 2013, \aap, 556, A2

\bibitem[{{Wang} {et~al}\mbox{.}(2013){Wang}, {Xu}, {An}, {Gu}, {Guo}, {Li},
  {Wang}, {Liu}, {Martineau-Huynh}, \& {Wu}}]{wang2013}
{Wang} J. {et~al.}, 2013, \apj, 763, 90

\bibitem[{{Watkinson} {et~al}\mbox{.}(2015){Watkinson}, {Mesinger},
  {Pritchard}, \& {Sobacchi}}]{watkinson15}
{Watkinson} C.~A., {Mesinger} A., {Pritchard} J.~R., {Sobacchi} E., 2015,
  \mnras, 449, 3202

\bibitem[{{Watkinson} \& {Pritchard}(2014)}]{watkinson14}
{Watkinson} C.~A., {Pritchard} J.~R., 2014, \mnras, 443, 3090

\bibitem[{{White} {et~al}\mbox{.}(2003){White}, {Becker}, {Fan}, \&
  {Strauss}}]{white03}
{White} R.~L., {Becker} R.~H., {Fan} X., {Strauss} M.~A., 2003, \aj, 126, 1

\bibitem[{{Yatawatta} {et~al}\mbox{.}(2013){Yatawatta}, {de Bruyn},
  {Brentjens}, {Labropoulos}, {Pandey}, {Kazemi}, {Zaroubi}, {Koopmans},
  {Offringa}, {Jeli{\'c}}, {Martinez Rubi}, {Veligatla}, {Wijnholds}, {Brouw},
  {Bernardi}, {Ciardi}, {Daiboo}, {Harker}, {Mellema}, {Schaye}, {Thomas},
  {Vedantham}, {Chapman}, {Abdalla}, {Alexov}, {Anderson}, {Avruch}, {Batejat},
  {Bell}, {Bell}, {Bentum}, {Best}, {Bonafede}, {Bregman}, {Breitling}, {van de
  Brink}, {Broderick}, {Br{\"u}ggen}, {Conway}, {de Gasperin}, {de Geus},
  {Duscha}, {Falcke}, {Fallows}, {Ferrari}, {Frieswijk}, {Garrett},
  {Griessmeier}, {Gunst}, {Hassall}, {Hessels}, {Hoeft}, {Iacobelli}, {Juette},
  {Karastergiou}, {Kondratiev}, {Kramer}, {Kuniyoshi}, {Kuper}, {van Leeuwen},
  {Maat}, {Mann}, {McKean}, {Mevius}, {Mol}, {Munk}, {Nijboer}, {Noordam},
  {Norden}, {Orru}, {Paas}, {Pandey-Pommier}, {Pizzo}, {Polatidis}, {Reich},
  {R{\"o}ttgering}, {Sluman}, {Smirnov}, {Stappers}, {Steinmetz}, {Tagger},
  {Tang}, {Tasse}, {ter Veen}, {Vermeulen}, {van Weeren}, {Wise}, {Wucknitz},
  \& {Zarka}}]{yatawatta13}
{Yatawatta} S. {et~al.}, 2013, \aap, 550, A136

\bibitem[{{Zahn} {et~al}\mbox{.}(2007){Zahn}, {Lidz}, {McQuinn}, {Dutta},
  {Hernquist}, {Zaldarriaga}, \& {Furlanetto}}]{zahn07}
{Zahn} O., {Lidz} A., {McQuinn} M., {Dutta} S., {Hernquist} L., {Zaldarriaga}
  M., {Furlanetto} S.~R., 2007, \apj, 654, 12

\end{thebibliography}

\appendix
\section{Behaviour of the quadrupole moment}
\label{sec:appendix_b}
\begin{figure}
\psfrag{xh1}[c][c][1][0]{{\LARGE $\xb$}}
\psfrag{Pcross}[c][c][1][0]{{\LARGE $r_{\rhohi,\rhoh}$}}
\psfrag{Proh1}[c][c][1][0]{{\LARGE $\overline{\delta T_b}^2k^3P_{\rhohi,\rhohi}/(2\pi^2)\, ({\rm mK^2})$}}
\psfrag{k = 0.122 Mpc-1}[c][c][1][0]{{\Large $k=0.12\,{\rm Mpc}^{-1}$}}
\psfrag{fidu}[c][c][1][0]{Fiducial$\,\,\,\,\,\,\,\,$}
\psfrag{fidu inh}[c][c][1][0]{{Clumping$\,\,\,\,$}}
\psfrag{xray=0.8}[c][c][1][0]{{UIB Dom$\,$}}
\psfrag{xrays=0.8}[c][c][1][0]{{SXR Dom}}
\psfrag{xraysh=0.5}[c][c][1][0]{{UV+SXR+UIB$\,\,\,\,\,\,\,\,\,\,\,$}}
\psfrag{pl=2.0}[c][c][1][0]{{PL 2.0}}
\psfrag{pl=3.0}[c][c][1][0]{{PL 3.0}}
\psfrag{den}[c][c][1][0]{{{\tiny $\overline{\delta T_b}^2k^3P_{\rhoh, \rhoh}/(2\pi^2)\,\,\,\,\,\,\,\,\,\,\,\,\,\,\,\,\,\,\,\,\,\,\,\,\,\,\,\,\,\,\,\,\,\,\,\,\,\,\,\,\,\,$}}}
\includegraphics[width=.5\textwidth,angle=0]{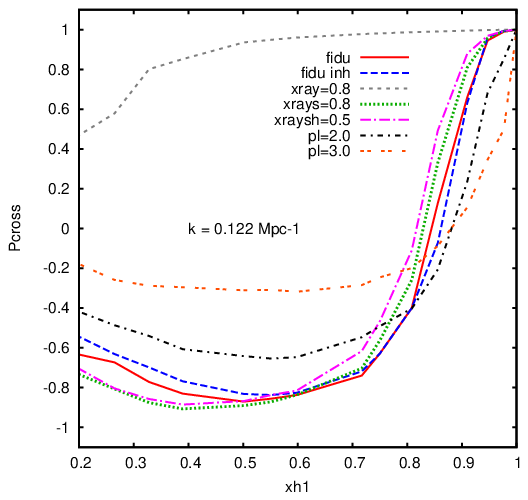}
\includegraphics[width=.5\textwidth,angle=0]{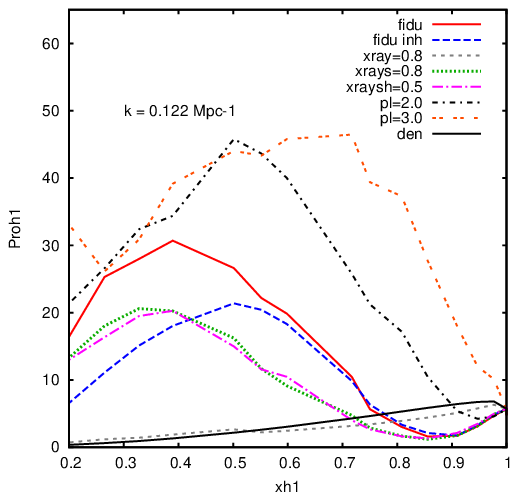}
\caption{The top and the bottom panels show the cross-correlation
  coefficient (or the cosine of the relative phase) between the matter
  and the \HI density fields, and the \HI density fluctuation power
  spectrum $P_{\rhohi,\rhohi}$, respectively, for all of our
  reionization scenarios, as a function of the global neutral fraction
  at $k=0.12$ Mpc$^{-1}$.}
\label{fig:p_cross_coef}
\end{figure}
The similarity in the behaviour of the quadrupole moment, $P_2^s$
(Figure \ref{fig:p2}), estimated for different reionization scenarios
may appear unintuitive. However, it can be explained using the
quasi-linear model (Equation \ref{eq:legendre_P2}), where the major
contribution in the evolution of $P_2^s$ has been ascribed to the
evolution of $P_{\rhohi,\rhoh}$ (Figure \ref{fig:p_cross}).

The cross-power spectrum between two fields, defined in Fourier space
as $X=Ae^{i\theta}$ and $Y=Be^{i\phi}$, will be $P_{XY} =
\operatorname{Re}\left(X^{\ast}Y\right) = |A||B| \cos(\phi-\theta)$.
Thus the cross-power spectrum between the matter and the \HI fields
($P_{\rhohi,\rhoh}$), is essentially a product of three quantities:
the amplitudes of the matter density fluctuations and the \HI density
fluctuations and the cosine of the phase difference between these two
fields. Since the matter density fluctuations are the same for all our
reionization scenarios, this means that the differences in
$P_{\rhohi,\rhoh}$ will be solely due to the differences in the
amplitude of their \HI density fluctuations and its relative phase
with the matter density field.

The top panel of Figure \ref{fig:p_cross_coef} shows the evolution of
the cosine of this phase difference for different reionization
scenarios estimated through the cross-correlation coefficient
$r_{\rhohi,\rhoh}(k) = P_{\rhohi,\rhoh}(k)/\sqrt{P_{\rhohi,\rhohi}(k)
  P_{\rhoh,\rhoh}(k)}$.  An obvious observation from Figure
\ref{fig:p_cross_coef}, is that the evolution of $r_{\rhohi,\rhoh}$ is
very similar in all of the reionization scenarios, except for the UIB
Dom scenario. This is because as long as the spatial distribution of
the major reionization sources follow the underlying matter
distribution, the relative phase between the \HI and the matter
density fields remains approximately the same and also evolves
similarly. In other words, $r_{\rhohi,\rhoh}$ evolves similarly as
long the reionization is inside-out in nature.

The bottom panel of Figure \ref{fig:p_cross_coef} shows the evolution
of the \HI density power spectrum $P_{\rhohi,\rhohi}$, which measures
the amplitude of fluctuations in the \HI field. The square root of
$P_{\rhohi,\rhohi}$ is the other quantity effectively contributing to
the evolution of $P_{\rhohi,\rhoh}$. It is evident from this figure
that the amplitude and the location of the peak in $P_{\rhohi,\rhohi}$
is significantly different for different reionization scenarios as it
is more susceptible to the \HI topology. In the cross-power spectrum,
however, it gets tuned by the phase difference ($r_{\rhohi,\rhoh}$)
term. This is what makes the contribution from $P_{\rhohi,\rhoh}$ in
$P_2^s$ similar in most of the reionization scenarios.

According to the quasi-linear model (Equation \ref{eq:legendre_P0}),
$P_0^s$ also contains a contribution from $P_{\rhohi,\rhoh}$, but in
addition to that it also contains contribution from the power spectrum
of the \HI field, $P_{\rhohi,\rhohi}$, which as described before,
depends only on the amplitude of the \HI fluctuations and not on its
relative phase with density field. The contribution from
$P_{\rhohi,\rhohi}$ makes the monopole moment $P_0^s$ more sensitive
to the \HI topology (or the \HII bubble size distribution) and thus
more dependent on the specific reionization scenario (a visual
comparison of Figure \ref{fig:p2} with the bottom panel of Figure
\ref{fig:p_cross_coef} further strengthens this argument). Because of
this the $P_2^s$ is a more robust measure of the reionization history than
$P_0^s$.

\end{document}